\pdfoutput=1
\documentclass[12pt,a4paper]{article}
\usepackage[english]{babel} % ,czech
\usepackage[utf8]{inputenc}
\usepackage[T1]{fontenc}
\usepackage{amsmath}
\usepackage{amsfonts}
\usepackage{textcomp}
\usepackage{graphicx}
\usepackage{mathrsfs}
\usepackage{amsbsy}
\usepackage{hyperref} %% pdftex option? 
\usepackage{cleveref}
\usepackage{xr}   %crossreferencing between two documents
\usepackage{grffile}
\usepackage{placeins}
\usepackage{ifthen}
\newboolean{arxiv} % 
\setboolean{arxiv}{true} 
\ifthenelse{\boolean{arxiv}}
{
}
{
\usepackage[style=ieee,maxcitenames=8,minbibnames=14,mincitenames=1,backend=biber,natbib=true]{biblatex} % maxbibnames=12,
}

\DeclareMathOperator{\di}{d\kern-0.4ex}

\newboolean{biblocal}
\setboolean{biblocal}{false}
\newboolean{supinfo} % 
\setboolean{supinfo}{true} 
\newboolean{mainonly} % 
\setboolean{mainonly}{false} 
\newboolean{figures} % 
\setboolean{figures}{true} 
\renewcommand{\columnwidth}{\textwidth}

\newcommand{\DPn}[3]{%
\frac{\partial^{#1}{#2}}
     {\partial{#3^{#1}}}
}

\newcommand{\DP}[2]{%
\frac{\partial{#1}}
     {\partial{#2}}
}

\ifthenelse{\boolean{arxiv}}
{
  \bibliographystyle{unsrt}
  \bibliography{plasmon-soliton-paper-V9-with-supplement-info_biber.bib}
}%else du test
{
  \defbibheading{bibliography}[\refname]{\centering{\textsc{References}:\vspace{-0.0cm}}}
  \ifthenelse{\boolean{biblocal}}
{
  \addbibresource{/home/renverse/LATEX/Biblio/Biblio-loe/bibliographie-articles-importes-wiley.bib}
  \addbibresource{/home/renverse/LATEX/Biblio/Biblio-loe/bibliographie-articles-importes-OSA.bib}
  \addbibresource{/home/renverse/LATEX/Recherches/Plasmons-solitons/Ex-Optica-2016/HMM-bibfiles-articles/Nonlinear-metamaterial-ENZ-v2.bib} 
  \addbibresource{/home/renverse/LATEX/Recherches/Plasmons-solitons/Ex-Optica-2016/HMM-bibfiles-articles/Linear-Hyperbolic-metamaterials-waveguides.bib} 
  \addbibresource{/home/renverse/LATEX/Biblio/Biblio-loe/bibliographie-articles-loe.bib}
  \addbibresource{/home/renverse/LATEX/Biblio/Biblio-loe/bibliographie-perso-articles-tri-an.bib}
  \addbibresource{/home/renverse/LATEX/Biblio/Biblio-loe/bibliographie-articles-importes-wiley.bib}
  \addbibresource{/home/renverse/LATEX/Biblio/Biblio-loe/bibliographie-articles-importes-OSA.bib}
  \addbibresource{/home/renverse/LATEX/Biblio/Biblio-loe/bibliographie-livres-loe.bib}
  \addbibresource{/home/renverse/LATEX/Biblio/Biblio-loe/bibliographie-articles-importes-AIP.bib}
  \addbibresource{/home/renverse/LATEX/Biblio/Biblio-loe/bibliographie-articles-importes-APS.bib}
  \addbibresource{/home/renverse/LATEX/Biblio/Biblio-loe/bibliographie-articles-importes-IOP.bib}
  \addbibresource{/home/renverse/LATEX/Biblio/Biblio-loe/bibliographie-articles-importes-nature-group.bib}
} %% else du test
{
  \addbibresource{plasmon-soliton-paper-V9-with-supplement-info_biber.bib} %% to generate it locally: biber --output-format=bibtex plasmon-soliton-paper-nature-photonics-V9-with-supplement-info.bcf 
} 
}

\hypersetup{pdfauthor={Gilles Renversez, Mathieu Chauvet},pdftitle={Plasmon--soliton waves: experimental demonstration and numerical simulations},pdfkeywords={spatial soliton, plasmon, Kerr nonlinearity, planar waveguide, chalcogenide glass},pdfsubject={research article, optics, nonlinear waves, experiments, modelling}}

\begin{document}
\title{Experimental demonstration of plasmon-soliton waves}

\author{Tintu Kuriakose$^1$, Gilles Renversez$^{2,*}$, Virginie Nazabal$^{3,5}$,  Mahmoud M. R.  Elsawy$^2$,\\Nathalie Coulon$^4$, Petr N\v{e}mec$^5$, and Mathieu Chauvet$^{1,*}$}
\maketitle
\noindent $^{1}$ FEMTO-ST Institute, CNRS, University of Bourgogne Franche-Comté, 15B avenue des Montboucons, 25030 Besançon, France\\
$^2$ Aix--Marseille Univ, CNRS, Centrale Marseille, Institut Fresnel, 13013 Marseille, France\\
$^3$ Univ Rennes, CNRS, ISCR (Institut des Sciences Chimiques de Rennes) -- UMR 6226, 35000 Rennes, France\\
$^4$ Univ Rennes, CNRS, IETR (Institut d'Electronique et de Télécommunications de Rennes) -- UMR 6164, 35000 Rennes, France\\
$^5$ Department of Graphic Arts and Photophysics, Faculty of chemical Technology, University of Pardubice, Studentská 573, 53210 Pardubice, Czech Republic\\
\vspace{0.15cm} $\,$\\
$^*$\texttt{\small mathieu.chauvet@univ-fcomte.fr} and \texttt{\small gilles.renversez@univ-amu.fr}\\

\begin{abstract}

Controlling low power light beam self-confinement with ultrafast response time opens up opportunities for the development of signal processing in microdevices. The combination of highly nonlinear medium with the tight confinement of plasmonic waves offers a viable but challenging configuration to reach this goal.  Here, we report the experimental observation of plasmon-soliton waves propagating in a chalcogenide-based four-layer planar geometry engineered to limit plasmon propagation losses while exhibiting efficient Kerr self-focusing at moderate power. The observations reveal a strongly enhanced self-focusing undergone by a self-trapped beam propagating inside the structure. As expected from theory, only TM polarized waves exhibit such a behaviour. Different experimental arrangements are tested that unambiguously reveal the nonlinear plasmon-soliton waves which are corroborated by simulations.\\
\vspace{0.5cm} $\,$\\
%\noindent OCIS code: (190.0190) Nonlinear optics, (250.5403) Plasmonics, (240.6680) Surface plasmons, (190.3270) Kerr effect, (190.6135) Spatial solitons, (230.7400) Waveguides, slab
\end{abstract}

\newpage
\selectlanguage{english}
The efficiency of nonlinear optical processes is well
known to benefit from extreme light confinement. One 
way to achieve such a strong electromagnetic field is to
exploit light interactions with metal nanostructures. The
surface plasmon polariton (SPP)~[1] at a metal-dielectric
interface, associated with both a collective oscillation of
free charges in the metal and an extremely confined
electromagnetic wave, has been a key discovery. Since
then, plasmonics has evolved into a flourishing research
field~[2]. It was then natural to combine the fields of plasmonics and nonlinear
optics to exploit these intense optical fields, allowing researchers to envision a variety of fascinating and
original physical phenomena with great application potential.
In addition to providing enhanced nonlinear effects with
ultra-fast response times, plasmonic nanostructures allow
nonlinear optical components to be scaled down in size.
As a consequence, the field of nonlinear plasmonics~[3] has grown significantly in recent years with
applications such as frequency conversion~[4],
switching and modulation~[5]. However, this field of
research is still in its infancy.

More specifically, first descriptions of one-dimensional
nonlinear plasmon-solitons and nonlinear surface waves
at metal/dielectric and dielectric/dielectric interfaces
were unveiled in the early eighties~[6–9]. The concept
being to propagate a plasmon-polariton wave at a metal-
nonlinear material interface in order to induce an exalted
Kerr self-focusing effect giving rise to a self-trapped
wave. Up to now, the problem has been tackled only
theoretically, based on analytical and numerical
calculations~[10–20] dealing with both spatial and
temporal beam trapping. Despite this constant and
comprehensive modelling development, no
experimental evidence of self-trapped nonlinear
plasmonic waves, named plasmon-solitons in~[10], has yet
been revealed despite almost four decades since the first
publication on nonlinear surface waves~[6-7]. Among the
practical challenges inherent to this plasmon assisted
self-focusing demonstration are the large propagation
losses associated with plasmons, the request for a strong
Kerr coefficient, and the limitations due to the damage
threshold intensity of the plasmonic structure.

In the present work, a four-layer slab component is
designed and fabricated to support a SPP with moderate
losses along with a strong Kerr nonlinear effect. This
structure allows the first experimental observation of a
hybrid plasmon-soliton wave that combines a spatial
soliton and a SPP in a single wave. It reveals a strong
enhancement of the beam two-dimensional self-trapping
efficiency due to the plasmonic effect occurring for such
type of nonlinear wave.
The content of the article is as follows: we start by
describing the structure design and fabrication steps.
Secondly, nonlinear optical characterizations are
presented. Finally, numerical simulations are confronted
to the experiments and results are discussed.

\begin{figure*}[!bt]
\centering
\ifthenelse{\boolean{figures}}{%
\includegraphics[width=1\textwidth,clip=true,trim= 0 0 0 0]{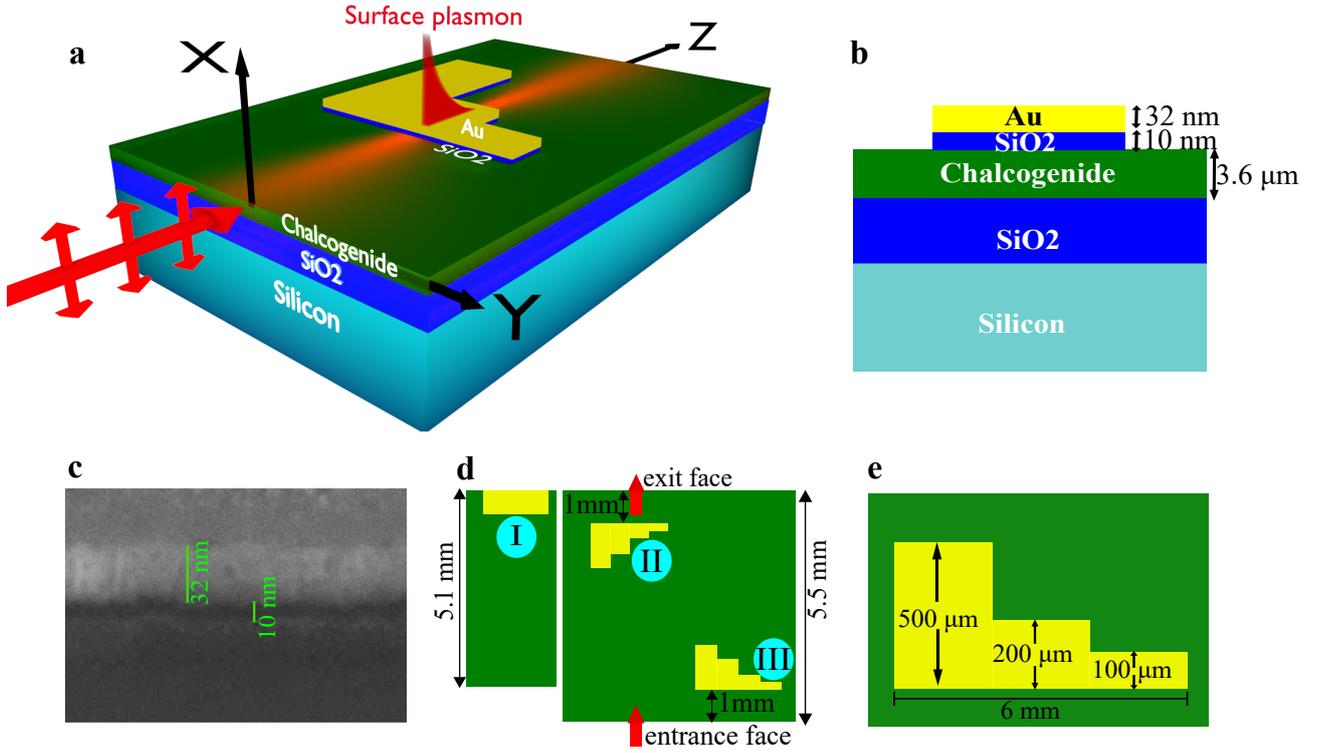}
}{}
\caption{View of the structure. a, artistic 3-D view of the experiment. b, schematic description of the cross section of the sample with the plasmonic structure. c, SEM image of the Au/SiO$_2$/ chalcogenide interfaces. d, schematic overall top view of the sample showing the three studied configurations denoted by I, II and III. e, detailed schematic top view of the used staircase metallic pattern.}
\label{fig:geometry-general-scheme}
\end{figure*}

\section*{Design and fabrication of the structures}

The origin of our nonlinear plasmonic component is a
planar chalcogenide waveguide that possesses the
suitable nonlinear properties to form Kerr spatial solitons
at near-infrared wavelengths, as demonstrated in refs~[21,22]. A metal film must be present for the
propagating wave to have a plasmonic part. However, a
compromise has to be found between the tight light
confinement provided by the plasmonic effect, which we
want to exploit to efficiently exalt the Kerr effect, and the
propagation losses induced by the metal that can be too
high and prevent the observation of the Kerr self-focusing. The chosen solution is to insert a thin silica
buffer layer between the metal and the nonlinear
dielectric layer. The designed waveguide including its
plasmonic structure (PS) derives from our previously
published theoretical studies~[13,14] and from Finite
Element Method simulations (Methods).
To be more specific, the Kerr layer is constituted of a 3.6
$\mu$m thick Ge$_{28.1}$ Sb$_{6.3}$Se$_{65.6}$ amorphous chalcogenide thin film~[23]
(Fig.~1 a-b and Methods). This composition was
chosen due to its large Kerr nonlinearity ($n_2$ = 5.5 10$^{-18}$ m$^2$/W), moderate two-photon absorption coefficient ($\alpha_2$ =
0.43 cm/GW) at the wavelength of interest of 1.55 $\mu$m,
and a high damage threshold intensity evaluated to be at 
approximately 2.5 GW/cm$^2$~[22,24]. In addition, this film
composition presents a reduced photo-sensitivity~[24]. The silica
buffer layer followed by a gold layer are then deposited
on top. Finite Element Method (FEM) numerical
simulations~[25, 26] have been performed to compute the
nonlinear modes propagating in the structure in order to 
determine the optimum thicknesses for the top layers. These simulations 
showed that the silica buffer layer thickness $d$ is the most
critical parameter for the design, as expected from a
previous study~[27].
 
As shown in Fig.~2a for d=30 nm, the fundamental optical
waves propagating in the structure are similar for TM
and TE polarizations and resemble the modes present
without metal. To the contrary, for d= 10 nm, the TM
mode clearly benefits from a strong localization due to
the plasmonic effect while the TE mode is nearly
unaffected (Fig.~2 and Methods). In addition, simulations
show that a 10 nm buffer layer allows a decrease by a
factor three of the propagation losses for the TM mode
compared to a basic structure without silica buffer (see
Supplementary information). Note that the deposition of
a very thin SiO$_2$ layer of good quality is challenging with
the operated sputtering deposition technique. To finalize
the structure, a 32 nm thick gold layer was then sputtered
on top, as illustrated in Fig.~1b. A cross section of the
fabricated structure was analyzed by SEM (Fig.~1c), 
confirming that the targeted and fabricated nanolayer
thicknesses are in accordance.
\begin{figure*}[!bt]
\centering
\ifthenelse{\boolean{figures}}{%
\includegraphics[width=0.875\columnwidth,clip=true,trim= 0 0 0 0]{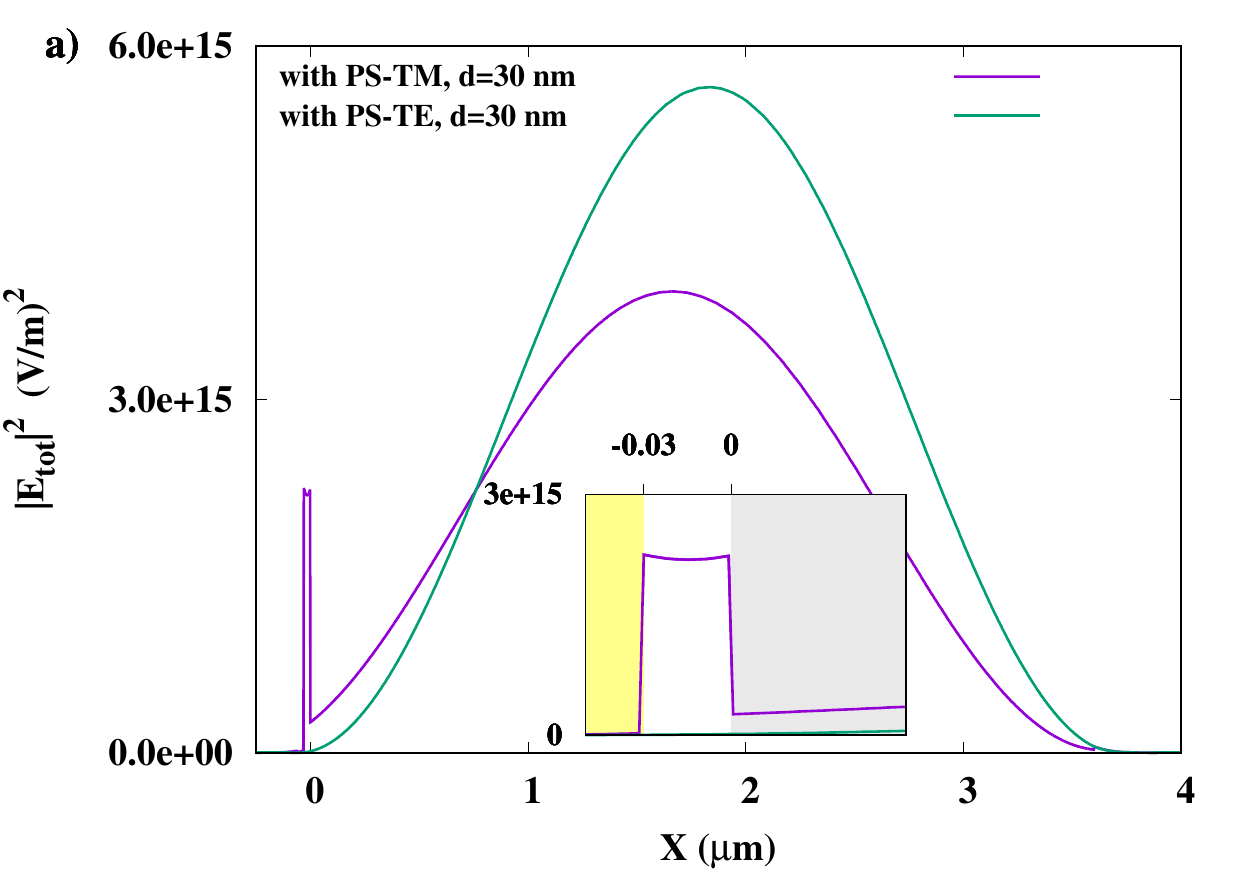}\\
\includegraphics[width=0.875\columnwidth,clip=true,trim= 0 0 0 0]{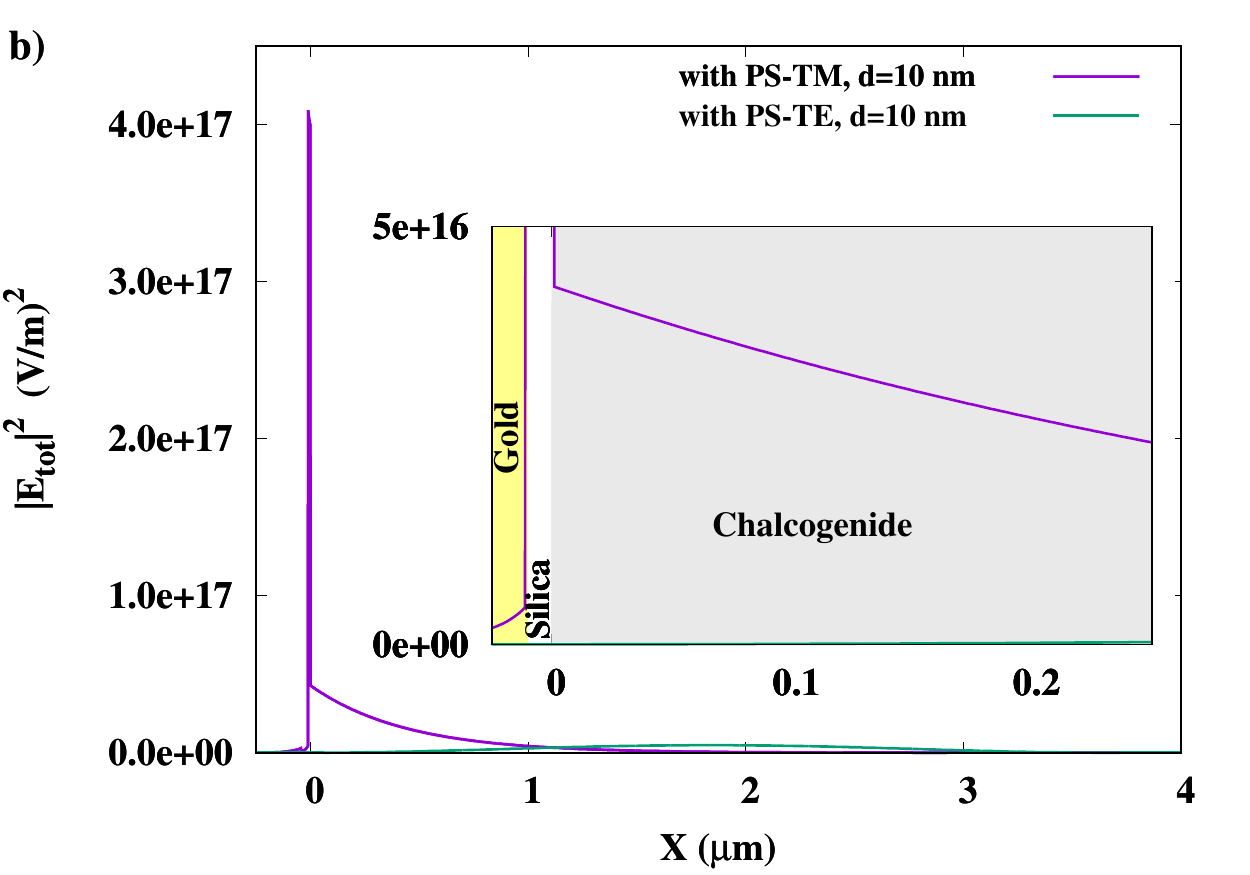}
}{}
\caption{Calculated mode intensity profiles along the X-axis and zoom-in near the
silica/gold interface(insets) of same power fundamental TM (green curves) and TE
(magenta curves) for SiO$_2$ buffer thickness $d$ equal to 30 nm (a) and 10 nm (b).}
\label{fig:field-profiles-TE-TM-h10nm-h30nm}
\end{figure*}

As depicted in Fig.~1d, the gold layer was patterned to
obtain either rectangular or staircase (Fig.~1e) PS area.
This arrangement provides the versatility to analyze
several configurations by choosing the position of
the launched beam along the Y-axis. For instance,
propagation was tested with different PS lengths $h$ and
different PS locations. Specifically, three different PS
locations were considered as described in Fig.~1d. The first
one was a rectangular plasmonic pattern, $h$ = 660  $\mu$m long,
located at the exit face of the 5.1 mm long sample
(configuration I). The two other configurations
corresponded  to a staircase PS with $h$ varying from 100  $\mu$m
to 500  $\mu$m, present either 1 mm before the exit face or
1mm after the entrance face to form configurations II and
III, respectively from a 5.5~mm long sample. The input and
output faces were formed by cleaving the processed wafer.

\section*{Optical characterizations}

Optical characterizations were performed with a 1550 nm
source emitting 200 fs pulses at a repetition rate of 80
MHz (Fig.~3a and Methods). The laser beam is
elliptically shaped to form a 4 $\mu$m by 31  $\mu$m (Fig.~3a)
beam waist (FWHM) along the X and Y-axes,
respectively. It is end-fire coupled into the waveguide so
that the 31 $\mu$m beam waist is located 2 mm inside the
sample. Such an arrangement lowers the intensity at the
entrance face compared to the intensity inside the
waveguide. It thus helps to prevent input facet damage in the
high intensity regime. In addition, it gives a weakly
diffracting beam over the 5.1 mm long propagation
distance. Careful beam alignment is also performed to
maximize the beam overlap with the fundamental mode
of the planar waveguide and thus avoid excitation of
higher-order modes. Observation at the output face of the
structure with a camera confirms that the presence of higher order modes
 is negligible. After the
propagation inside the
waveguide, the diffracted beam reaches a  FWHM of 41 $\mu$m
along the Y-axis (Fig.~3a) in the linear regime (low intensity). In
this regime a $ \sim$ 21\% waveguide transmission is measured
for both TM and TE polarizations when light propagates
away from the metallized area. This transmitted power is
consistent with a coupling efficiency of 28\% and
propagation losses of about 0.19 cm$^{-1}$ .

\begin{figure*}[!bt]
\centering
\ifthenelse{\boolean{figures}}{%
\includegraphics[width=1\textwidth,clip=true,trim= 0 0 4 0]{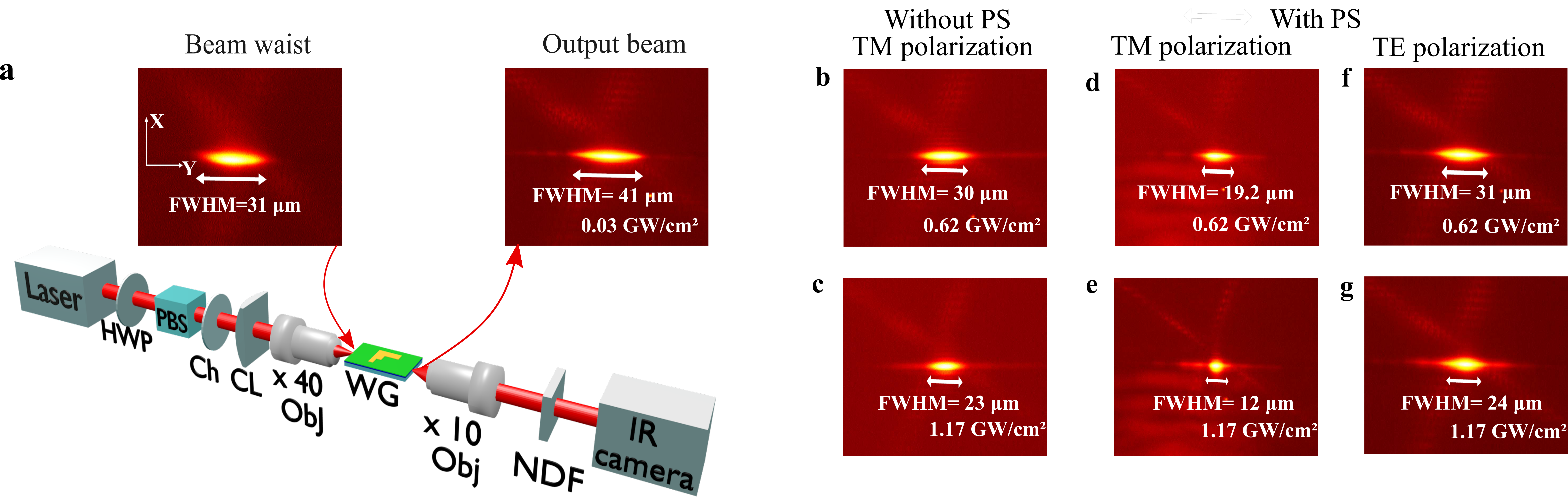}
}{}
\caption{a, Experimental setup used for the observation of the plasmon-soliton wave and intensity distribution of the beam waist and output beam in linear regime. HWP; half wave plate, PBS; polarizing beam splitter, Ch; chopper, CL; cylindrical lens, Obj; microscope objective, NDF; neutral density filters. b--g, Intensity distribution of self-focused output beams for two input intensities without PS (b,c), with PS corresponding to configuration I for TM polarization (d,e) and for TE polarization (f,g).}
\label{fig:setup-output-field-images}
\end{figure*}

To unambiguously reveal the influence of the plasmon-soliton wave on the beam self-focusing, we first
characterize the self-trapping behaviour in the absence of
metallic layer along the beam path. These preliminary
experiments can thus be considered as a reference. As
shown in references [21,22] the optical Kerr nonlinearity
in similar chalcogenide slab waveguides can support
spatial solitons at near-infrared wavelengths. In the tested
sample we observe that, for an input intensity of 0.62
GW/cm$^2$ , a 30 $\mu$m FWHM beam is obtained at the output
face (Fig.~3b). This size is close to the injected beam
waist of 31 $\mu$m (Fig.~3a) which indicates that diffraction
is compensated by the nonlinear Kerr self-focusing effect
to form a spatial soliton [28].

A stronger focusing effect occurs If the input intensity is  raised  further. For an intensity of 1.17 GW/cm$^2$ a 23 $\mu$m
FWHM is obtained at the output (Fig.~3c). Note that the
very same behaviour is observed for both TM and TE
polarizations for this configuration without PS.
We then shift laterally the sample so that a PS is present
on the trajectory of the launched beam. To be more
specific, configuration I described in Fig.~1d is first
considered with a PS length $h$ of 660  $\mu$m. In the linear
regime the presence of this metallic structure located near
the end of the waveguide does not modify the observed
output beam distribution. The linear transmission is reduced
however  due to additional propagation losses
compared to the situation without the PS. We deduce that
under the PS the attenuation is of 0.57 cm$^{-1}$ and 28 cm$^{-1}$ for
TE and TM polarized light, respectively. The large
attenuation for the TM case is a first indication of the
plasmonic part of the plasmon-soliton predicted in Fig.~2b. Subsequently, as the intensity is raised, a strong enhancement of the self-focusing behaviour is observed at
the output of the waveguide for TM waves. For instance,
for an input intensity of 0.62 GW/cm$^2$ (Fig.~3d) the
output beam is already narrower (19 $\mu$m FWHM) than
the injected one (31 $\mu$m FWHM) while an even higher
input intensity of 1.17 GW/cm$^2$ leads to a very efficient
trapping of the beam, as witnessed by the output beam of
12 $\mu$m FWHM (Fig.~3e). By comparing this result with
the one obtained at the same intensity without PS (Fig.
3c) we can deduce that the plasmonic enhanced nonlinear
effect induced by the 660 $\mu$m long PS focuses the beam
from about 23 $\mu$m to 12 $\mu$m FWHM . It is important to
note that the exalted focusing is not observed for the TE
polarization. Indeed, as shown in Fig.~3f-g, the TE wave
shows no improved self-focusing compared to the
configuration without the PS (Fig.~3 b-c). This
polarization dependent behavior is a characteristic feature
of the plasmonic effect on the field profile. To exclude
the possible influence of thermally-induced self-focusing the experiment was repeated by inserting a
mechanical chopper after the light source to diminish 
the average power to 40\% while keeping the same peak
power. We did not notice a measurable change compared to the behaviour described in Fig.~3 that excludes any
significant role of the temperature (see Section 6 in
Supplementary information).

\begin{figure*}[!bt]
\centering
\ifthenelse{\boolean{figures}}{%
\includegraphics[width=1\columnwidth,clip=true,trim= 0 0 0 0]{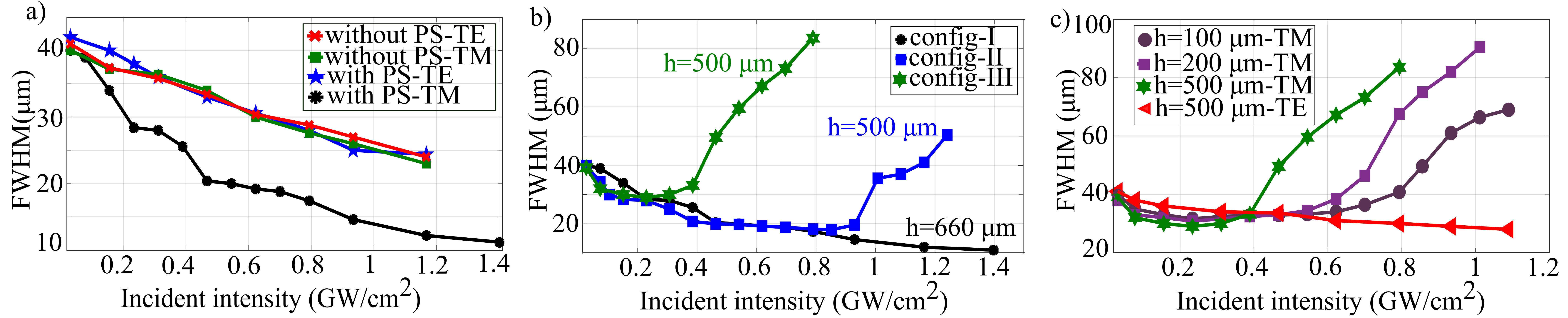}
}{}
\caption{a, Measured output beam FWHM as a function of input intensity for both TM and TE polarizations without PS and with a  PS  corresponding  to  configuration  I.  b,  Evolution  of  output  beam FWHM  as  a  function  of  input  intensity  for configurations I, II and III for TM polarization. c, Evolution of output beam FWHM for configuration III for PS lengths $h$ of 500
   $\mu$m, 200 $\mu$m and 100 $\mu$m for TM polarization and for $h$ of 500 $\mu$m for TE polarization. }
\label{fig: }
\end{figure*}

The experimentally demonstrated striking self-focusing
enhancement due to the plasmonic effect is also evident
in the graph in Fig.~4a where the measured output
FWHM beam is plotted as a function of the input beam
intensity for TE and TM polarizations, both with and
without the PS. For the three arrangements for which
plasmon generation is not achievable, i.e. without or
with PS for TE polarization, a similar behaviour is
observed. This behaviour is characterized by a gradual focusing with a
linear evolution of the beam width versus intensity. For a TM polarized beam in presence of a PS, it 
differs significantly in this configuration I with h=660
 $\mu$m. 1) We observe that this plasmon-enhanced
configuration always gives rise to an output FWHM
beam smaller than for the other three standard
configurations. 2) The focusing effect takes place at a
faster pace for low intensity. 3) Three times less intensity
is required compared to the case without PS enhancement
to obtain an output beam of similar size than the coupled
beam waist of 31 $\mu$m. 4) The FWHM evolves with a
nonlinear trend and tends to saturate to a beam size
approaching a narrow 11 $\mu$m FWHM value at high
intensity (1.4 GW/cm$^2$).
Indeed, due to the linear and two photon absorptions, the
intensity reaching the PS is not reduced as much as in
configuration III where the PS is far from the input face.

To gain a better understanding of the disclosed plasmon-soliton wave that propagates within the PS and of the
associated strongly enhanced focusing, additional
configurations were tested. We first studied the influence
of the position of the PS along propagation. Figure.~4b
compares the observed FWHM as a function of intensity
for a 500 $\mu$m long PS positioned either 1 mm before the
exit face (Fig.~1d config. II) or 1 mm after the input face
(Fig.~1d config. III) and the previously described case of
the 660 $\mu$m long PS located at the exit face. At very low
intensity the beam undergoes the same free diffraction in
the 5.5 mm long sample regardless of the PS location, precluding any linear beam distortion from the PS.
As soon as the intensity is raised the output beam 
focuses sharply in accordance with the Kerr-lens effect present in
the PS. However, when the intensity is increased above
an intensity threshold a spreading of the beam is
observed for a PS located before the output face. The
threshold is about 0.35 GW/cm$^2$ and 0.87 GW/cm$^2$ for a
PS positioned 1 mm after the entrance face (config. III) or
1 mm before the exit face (config. II), respectively. This
behaviour can once again be explained by the strong
localized self-focusing induced by the PS. Indeed, when
such a focusing leads to a narrow spot the beam
diffraction cannot be compensated anymore by the
weaker nonlinearity present after the PS and the beam
diffraction is dominant. Consequently, beam spreading is
observed at the output face. The diffraction is weaker in
configuration II since the PS is only 1 mm from the
output observation face. Note that, as observed
experimentally, the input threshold intensity is expected
to be weaker when the PS is close to the entrance face.
Indeed, due to the linear and two photon absorptions, the
intensity reaching the PS is not reduced as much as in
configuration III where the PS is far from the input face.

\begin{figure*}[!bt]
\centering
\ifthenelse{\boolean{figures}}{%
\includegraphics[width=1\columnwidth,clip=true,trim= 0 0 0 0]{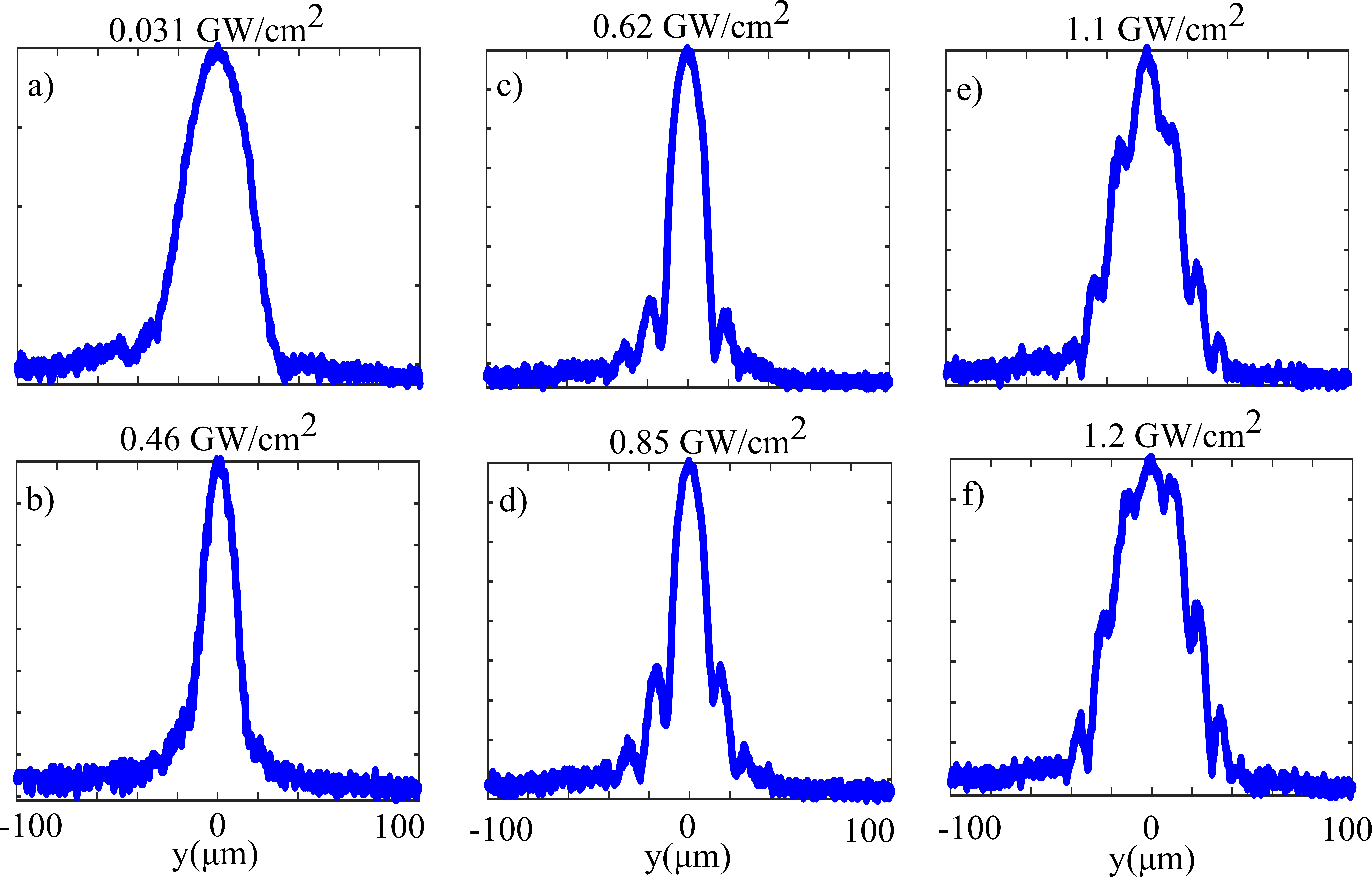}
}{}
\caption{Output beam profile evolution as a function of
incident intensity for configuration II, TM polarization,
h=500 $\mu$m.}
\label{fig: }
\end{figure*}

To analyze the impact of the length $h$ of the PS on the
enhanced self-focusing observed for TM waves,
experiments were conducted for PS lengths of
100 $\mu$m and 200 $\mu$m for configuration III (Fig.~1d-e). The
results are presented in Fig.~4c along with measurements
made for $h$ = 500 $\mu$m and for the enhancement free TE
polarization. As the intensity increases, the PS self-focusing enhancement focuses the beam to a minimum
value before a spreading of the FWHM occurs. The
shorter the PS, the weaker  the nonlinear focusing
effect. The minimum beam size is reached at an intensity
of about 0.24 GW/cm$^2$, giving an output FWHM of 31.6
 $\mu$m, 30.7 $\mu$m and 29 $\mu$m for 100 $\mu$m, 200 $\mu$m and 500
 $\mu$m PS length, respectively. As the intensity is
increased  further, the defocusing occurs with a reverse 
dependency with the PS length: the shorter the PS
length, the weaker the defocusing. This hints that even a PS as short as 100 $\mu$m can induce strong
focusing.

An analysis of the output beam profile provides crucial
additional information on the strong nonlinearity seen by
the propagating beam that was not revealed by the
FWHM measurements. In Fig.~5 the output beam profile
is plotted for six intensity values for a 500 $\mu$m long PS
positioned 1 mm from the output face (config. II). The
 efficient focusing effect  described previously forms at first 
 a smooth profile beam with a weak pedestal (Fig.~5b). This
beam then enlarges and develops symmetric side lobes
on both sides of the main peak as the intensity is raised
(Fig.~5). This behaviour along with the 
plasmon enhanced self-focusing described  above are corroborated by
numerical calculations presented in the following section.

\section*{Simulations-Discussion}
\label{sec:simulations-discussion}

In order to confirm the plasmon-soliton propagation in
the PS and to gain insight into the observed phenomena,
we built a numerical scalar model based on a modified
spatial nonlinear Schrödinger equation (SNLSE). This model is
adapted to take into account the mode field profile along
 the X-axis in the different sections of the full structure
(Methods). Without PS, the field profiles of both TE and
TM modes are nearly identical (Fig.~2a), and thus are the associated modal
parameters. In contast, in
presence of a PS with a 10 nm thick SiO$_2$ layer, the
propagation constants are slightly different (typically
below 2\% in the studied power range), but more importantly the field profiles are radically
different. This dissimilarity gives at least a threefold
enhancement of the effective nonlinearity parameter for
the TM mode compared to the TE mode (See
Supplementary information).

Fig.~\ref{fig:numerical-simulations}a presents the calculated profile evolution of the
beam intensity from the SNLSE along the Y-axis versus
propagation inside the full structure for the TM
polarization for configuration II with $h$ = 500 $\mu$m at an
input intensity (incident intensity I=1.25 GW/cm$^2$ ) with
the best numerically chosen enhancement factor for the
effective nonlinearity (eight instead of three, see Supplementary
information) that corresponds to the experiment depicted in
Fig.~5f. This emblematic case reveals several remarkable
characteristics that give more insight into the behaviour
unveiled experimentally. First, we clearly see that the
beam self-focuses along the first 2 mm in the structure and 
then it becomes nearly invariant before converging
dramatically to a highly focused central peak as it travels
in the 500 $\mu$m long PS region limited by the dashed lines.
Finally, as the beam leaves this highly nonlinear PS
region, diffraction suppasses the self-focusing effect. It
is important to note that the numerical model also predicts
the symmetric lateral peaks observed experimentally in
Fig.~5. These peaks are generated in the last part of the
PS for intensities above 0.6 GW/cm$^2$,  in fair
agreement with the experiments. Spatial modulation
instability can be at the origin of the beam break-up of an
initial noisy broad beam propagating in Kerr nonlinear
media but it does not fit to the present phenomena.
Indeed, the measured intensity dependence of the
generated spatial frequency does not follow the scaling
law of modulation instability~[29,30]. Moreover, large
amplitude spatial noise added in the present simulations
does not give rise to the observed multi-peaks unless the
plasmon-enhanced nonlinearity is present. We think the  phenomenon
 described here is a combination of the
beam break-up described in references [31,32] induced at the interface between a dielectric region with low
Kerr-type nonlinearity and one with a high
nonlinearity one and of the symmetric lateral peaks generated in overpowered
spatial solitons in nonlinear full vector simulations (see Supplementary information).

\begin{figure*}[!bt]
\centering
\ifthenelse{\boolean{figures}}{%
\includegraphics[width=0.75\columnwidth,clip=true,trim= 0 0 0 0]{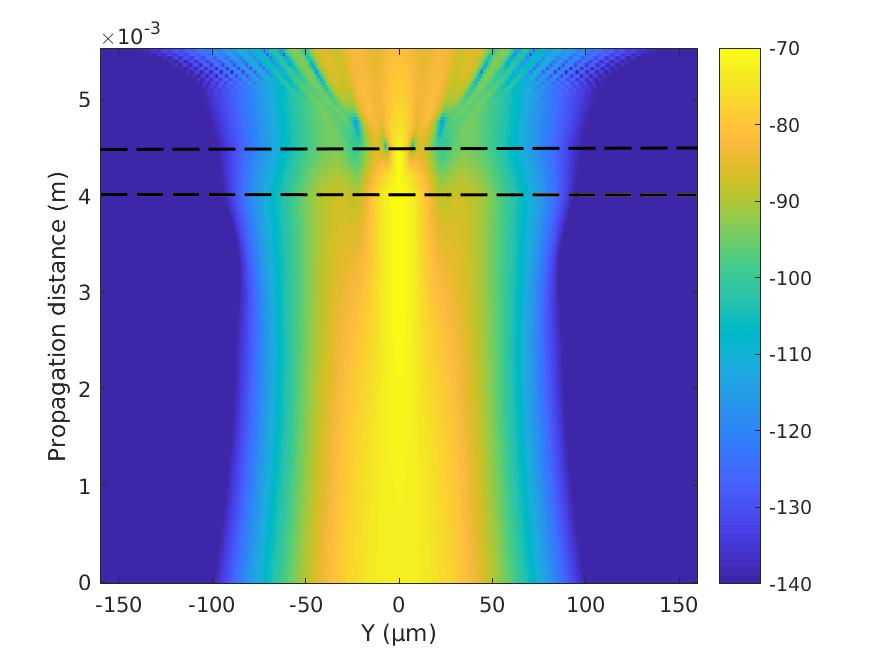}
\includegraphics[width=0.7\columnwidth,clip=true,trim= 0 0 0 0]{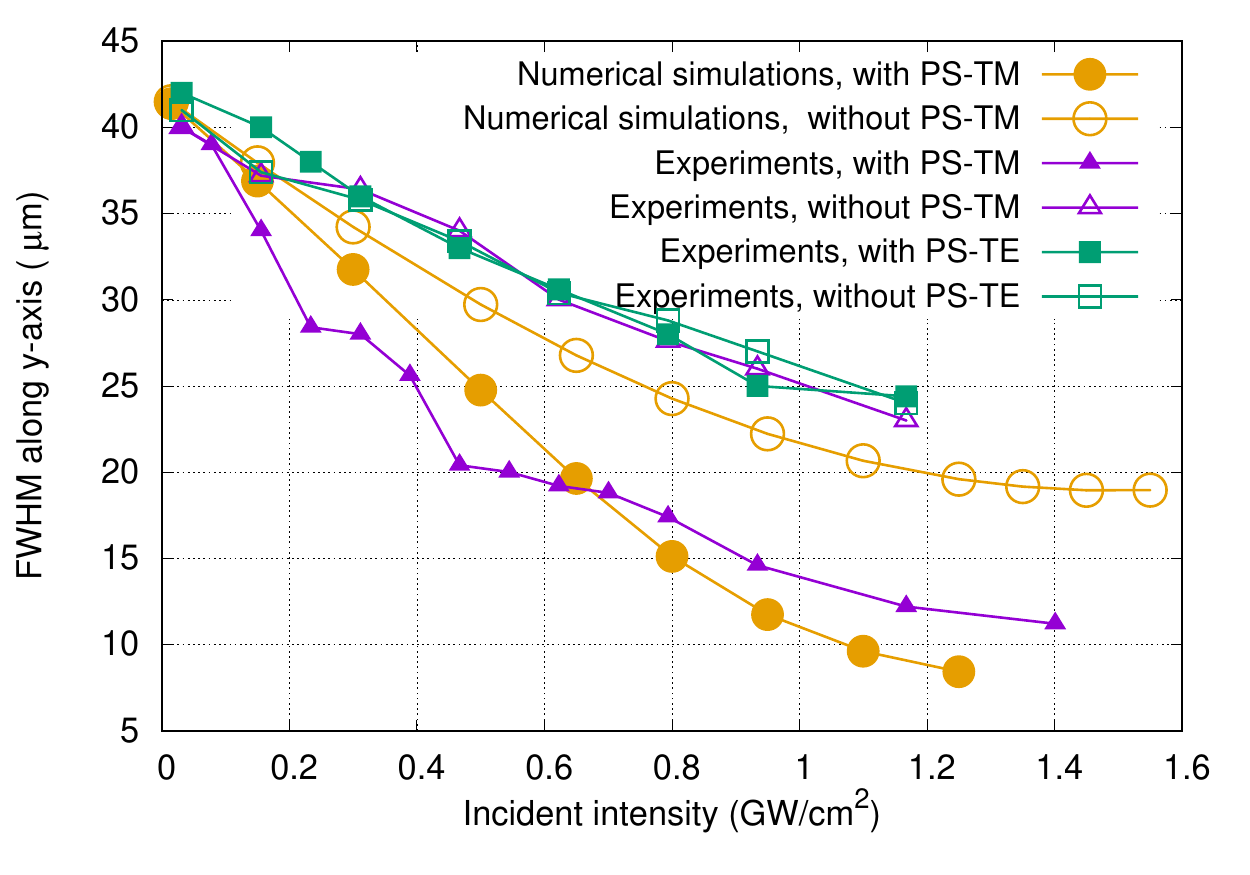}
}{}
\caption{Numerical results. a, color map in log scale of the beam intensity
evolution along the Y-axis versus propagation inside the
full structure for the TM polarization for configuration II
with $h$ = 500 $\mu$m in nonlinear regime (input incident
intensity of 1.25 GW/cm$^2$). The dashed lines represent
the limits of the PS. b, comparison of the computed and
measured FWHM for configuration I as a function of the
input beam intensity.}
\label{fig:numerical-simulations}
\end{figure*}

The saturation of the self-focusing effect appearing at
high power is also observed in the simulations.
Discrepancies between theory and experiments remain. First,
some limitations of the model can be invoked: it is based
on a scalar SNLSE while the true problem is a
vector one; paraxial approximations and calculation of
the effective nonlinearity are not fully valid when the beam
profile rapidly transforms in the PS; and the considered
instantaneous response of the Kerr effect could be
questioned in the femtosecond regime~[33] (see Supplementary information). Second, there is
some uncertainty on the values of the nonlinear
coefficients  $n_2$ and $\alpha_2$ of this uncommon chalcogenide composition~[34] (see also section~\ref{sec:nonlinear-prop-equation-derivation-model} in Supplementary information), and also on the exact values of the permittivity of the gold nanolayer and of its thickness along the plasmonic structure. Finally, potential nonlinear effects present in the gold layer~[35-37] can also be a reason of discrepancies between theory and measurements, especially in the high intensity regime.

\FloatBarrier
\section*{Conclusion}
\label{sec:conclusion}
In summary, a compelling spatial confinement of light
observed only for TM polarization in a Kerr plasmonic
structure arising from a plasmon-soliton wave has been
demonstrated experimentally. The designed structure
consists of a slab waveguide made of a highly nonlinear
chalcogenide layer covered with nanolayers of silica and
gold. The thickness of the silica layer has been tailored to
benefit from the enhanced self-focusing due to the high-intensity plasmon while limiting the detrimental effect of propagation losses induced by the metal. Experimental
observations are confirmed by numerical predictions
obtained using an improved model of nonlinear
propagation combining FEM modal results and spatial
nonlinear Schrödinger equation. This first experimental
proof validates the concept of the nonlinear spatial
optical self-trapping enhanced by the assistance of a
plasmon-soliton. It also allows envisioning the possibility of
extending the demonstration to other nonlinear surface
waves. In addition, the dramatic nonlinear beam
reshaping induced over a distance of few hundreds
microns at moderate light power also opens up new
perspectives for nonlinear integrated plasmonic devices.

\ifthenelse{\boolean{mainonly}}{%
}{%
\vspace{0.5cm}
\noindent \textbf{Acknowledgements}\\
The authors would like to acknowledge the financial
support of the Region Franche-Comt\'e and of Aix-Marseille University, and of the Czech Science
Foundation (project 16-17921S). This work was also
partially supported by the FEMTO-ST MIMENTO
technological facilities. We thank Tom\'aš Halenkovič for his contribution to the chalcogenide glass structure fabrication, Gwenn Ulliac for
taking SEM image of the sample,  and Xavier Romain for
the graphics help.

\noindent \textbf{Author contributions}\\
G.R. proposed the experiment. G.R., M.C. and T.K.
designed the PS structure. V.N., P. N. and N. C. fabricated
the samples. T.K. performed the optical characterizations.
G.R. and M.E. conducted the finite element method simulations, G.R. built  the nonlinear propagation model and conducted the associated simulations. T.K. assisted with the results analysis. M.C. and G.R. provided in-depth discussion of the project and analyzed the results. M.C., G.R. and T.K. wrote the manuscript.

\section*{Methods}
\label{sec:methods}

\noindent \textbf{Structure fabrication}\\
The heterostructure fabrication was realized in several steps. The chalcogenide glass target has been
specially manufactured to accommodate RF sputtering magnetron deposition. First, a  3.6 $\mu$m thick Ge$_{28.1}$Sb$_{6.3}$Se$_{65.6}$
 chalcogenide layer with a refractive index of 2.54 at 1.55  $\mu$m was deposited by RF
magnetron sputtering method on an oxidized silicon wafer. The morphology and topography of the
chalcogenide waveguides and heterostructure was characterized by SEM and AFM and showed no cracks
with an RMS roughness of about 0.3 nm for chalcogenide films. The refractive index dispersion was
determined by m-lines method and variable angle spectroscopic ellipsometry using Cody–Lorentz modelling. Then a 10 nm
silica buffer layer is deposited by the sputtering technique followed by a 32 nm gold evaporated layer
which is patterned by photolithography. The choice of a 32 nm thick gold nanolyer is a compromise between the need for a good quality layer with a constant thickness and the need to reduce the propagation losses generated by the metal knowing that the thinner the gold layer the smaller the losses. 

\noindent \textbf{Characterizations}\\
Optical measurements were performed with an
optical parametric oscillator emitting 200 fs pulses at a
repetition rate of 80 MHz at 1550 nm (Chameleon OPO from
Coherent). The power was varied by the combination of a half
wave plate and a polarizer. The latter component is also used to
adjust the polarization state of the injected beam to TM or TE
polarization. An optical chopper can also be inserted to change
the time average power. The laser beam is shaped into an elliptical
spot using a cylindrical lens and a x40 microscope objective to
form 4 $\mu$m and 31 $\mu$m beam waists (FWHM) along X-axis and Y-
axis, respectively. The beam was end-fire coupled into the
waveguide so that the 31 $\mu$m beam waist was located 2 mm inside
the sample. The beam profile at the output of the sample was 
imaged on a Vidicon camera with a x10 microscope objective.
Input and output powers were monitored by power meters to
deduce the coupling efficiency, the peak power and the
waveguide transmission. A coupling efficiency of 28\% and
propagation losses of about 0.19 cm$^{-1}$ are measured away from
PS structures. The input spatial averaged incident intensity was 
determined by dividing the coupled light peak power by the
elliptical section of the guided mode at the entrance face
corresponding to waists of 1.21 $\mu$m along the X-axis and 30 $\mu$m
(35.5 $\mu$m FWHM) along the Y-axis.

\noindent \textbf{Simulations}\\
\noindent Modal analysis\\
Finite Element Method (FEM) numerical simulations~[25,26]
 were performed to compute the nonlinear modes in the
structures with PS in order determine the optimum thicknesses
of the investigated layered structured. These simulations show
that the silica buffer thickness $d$ is the most critical parameter
for the design as expected from previous studies~[13,14,27]. 
As for all modal analysis, the investigated structure is assumed to
be invariant along the propagation direction in the modeling.

As depicted in Fig.~2a, for a silica buffer layer thickness $d$ = 30 nm, the TM and TE fundamental mode intensity profiles along
the X-axis at a given power weakly differ. The presence of the
metal marginally influences their distribution, as witnessed by
the calculated profiles which are very similar to the one of the fundamental
guided modes of the silica (thick bottom layer)/chalcogenide
slab waveguide. On the contrary, the TM mode profile is more
and more affected by the metal as the silica buffer layer becomes
thinner.
For TM polarization, the studied metal/dielectric structure (Fig.~1
b) supports a plasmon-soliton wave that extends into the nonlinear
dielectric layer. The propagation losses are weaker compared to
the case of the extremely confined plasmonic waves present at a
basic metal-dielectric interface. For our designed multilayer
structure we expect plasmon propagation over a few
hundreds micrometers. The plasmonic tail present in the
nonlinear layer efficiently activates the Kerr effect. This
configuration is thus favorable to reveal the formation of a self-focused beam that takes advantage of the plasmon-enhanced Kerr effect.
Guided by our previously published theoretical studies~[13,14,27], the FEM simulations provide an optimal value, in
terms of nonlinearity enhancement and propagation losses, of 10
nm for the thickness of the silica buffer layer.
It is worth mentioning that since the X and Y field profiles of the
2D plasmon-soliton nonlinear wave are coupled, the observed
localization along the X axis (Fig.~2 b, TM polarization) is
associated with an enhanced localization along the Y axis (See full
profile view in Section 1 of the Supplementary information).

\noindent Propagation simulations\\
To study the propagation of the input beam along the full
structure, as described in Fig.~1, which is no longer invariant
along the Z direction we built a numerical model based on the
spatial nonlinear Schrödinger equation (SNLSE) adapted to take
into account the mode field profiles along X-axis in the full structure. This scalar model
describes the transverse field profile along the Y-axis and its
evolution versus propagation along the Z-axis~[28]. The three
different configurations (I, II, or III) including the different
lengths $h$ of the PS can be studied. The field dependency along
the vertical X-axis does not explicitly appear in the model but is
considered indirectly through the Z-dependent modal properties
(effective nonlinearity and propagation constants) that appear in
the SNLSE. These modal parameters are obtained from the FEM-based simulations of the main TE and TM modes associated 
with the different sections of the full structure~[25,26].
Specifically, the effective nonlinearity is computed using several
integrals of the computed modal field~[26].

\section*{References}

\begin{enumerate}
\item S. L. Cunningham, A. A. Maradudin, and R. F.
Wallis, "Effect of a charge layer on the surface-
plasmon-polariton dispersion curve," Phys. Rev. B
10, 3342 (1974).
\item S. A. Maier, Plasmonics: Fundamentals and
Applications (Springer, 2007).
\item M. Kauranen and A. V. Zayats, "Nonlinear
plasmonics," Nat. Photonics 6, 737 (2012).
\item H. J. Simon, D. E. Mitchell, and J. G. Watson,
"Optical Second-Harmonic Generation with
Surface Plasmons in Silver Films," Phys. Rev.
Lett. 33, 1531–1534 (1974).
\item M. Z. Alam, J. S. Aitchison, and M. Mojahedi, "A
marriage of convenience: Hybridization of surface
plasmon and dielectric waveguide modes," Laser
Photonics Rev. 8, 394–408 (2014).
\item V. M. Agranovich, V. Babichenko, and V. Y.
Chernyak, "Nonlinear surface polaritons," JETP
Lett 32, 512–515 (1980).
\item W. J. Tomlinson, "Surface wave at a nonlinear
interface," Opt. Lett. 5, 323–325 (1980).
\item J. Ariyasu, C. T. Seaton, G. I. Stegeman, A. A.
Maradudin, and R. F. Wallis, "Nonlinear surface
polaritons guided by metal films," J. Appl. Phys.
58, 2460–2466 (1985).
\item N. N. Akhmediev, "Novel class of nonlinear
surface waves: asymmetric modes in a symmetric
layered structure," Sov Phys JETP 56, 299–303
(1982).
\item E. Feigenbaum and M. Orenstein, "Plasmon-
soliton," Opt. Lett. 32, 674–676 (2007).
\item A. R. Davoyan, I. V. Shadrivov, and Y. S. Kivshar,
"Self-focusing and spatial plasmon-polariton
solitons," Opt. Express 17, 21732–21737 (2009).
\item K. Y. Bliokh, Y. P. Bliokh, and A. Ferrando,
"Resonant plasmon-soliton interaction," Phys.
Rev. A 79, 041803 (2009).
\item W. Walasik, V. Nazabal, M. Chauvet, Y.
Kartashov, and G. Renversez, "Low-power
plasmon–soliton in realistic nonlinear planar
structures," Opt. Lett. 37, 4579–4581 (2012).
\item W. Walasik, G. Renversez, and Y. V. Kartashov,
"Stationary plasmon-soliton waves in metal-dielectric nonlinear planar structures: Modeling
and properties," Phys. Rev. A 89, 023816 (2014).
\item W. Walasik and G. Renversez, "Plasmon-soliton
waves in planar slot waveguides. I. Modeling,"
Phys. Rev. A 93, 013825 (2016).
\item W. Walasik, G. Renversez, and F. Ye, "Plasmon-
soliton waves in planar slot waveguides. II.
Results for stationary waves and stability
analysis," Phys. Rev. A 93, 013826 (2016).
\item A. Marini and F. Biancalana, "Ultrashort Self-
Induced Transparency Plasmon Solitons," Phys.
Rev. Lett. 110, 243901 (2013).
\item D. A. Smirnova, I. V. Shadrivov, A. I. Smirnov,
and Y. S. Kivshar, "Dissipative plasmon-solitons
in multilayer graphene", Laser Photonics Rev. 8,
No. 2, 291–296 (2014)
\item M. L. Nesterov, J. Bravo-Abad, A. Nikitin, F. J.
Garcia-Vidal, and L. Martin-Moreno, "Graphene
supports the propagation of subwavelength
opticals solitons", Laser Photonics Rev. 7, No. 2,
L7–L11 (2013)
\item A. Pusch, I. V. Shadrivov, O. Hess, and Y. S.
Kivshar, "Self-focusing of femtosecond surface
plasmon polaritons," Opt. Express 21, 1121–1127
(2013).
\item M. Chauvet, G. Fanjoux, K. P. Huy, V. Nazabal, F.
Charpentier, T. Billeton, G. Boudebs, M.
Cathelinaud, and S.-P. Gorza, "Kerr spatial
solitons in chalcogenide waveguides," Opt. Lett.
34, 1804–1806 (2009).
%\item T. Kuriakose, E. Baudet, T. Halenkovič, M. M. R. Elsawy, P. Němec, V. Nazabal, G. Renversez, and
\item T. Kuriakose, E. Baudet, T. Halenkovic, M. M. R. Elsawy, P. Nemec, V. Nazabal, G. Renversez, and
M. Chauvet, "Measurement of ultrafast optical
Kerr effect of Ge–Sb–Se chalcogenide slab
waveguides by the beam self-trapping technique,"
Opt. Commun. 403, 352–357 (2017).
\item V. Nazabal, F. Charpentier, J.-L. Adam, P. Nemec,
H. Lhermite, M.-L. Brandily-Anne, J. Charrier, J.-
P. Guin, and A. Moréac, "Sputtering and Pulsed
Laser Deposition for Near- and Mid-Infrared
Applications: A Comparative
Study of Ge$_{25}$Sb$_{10}$S$_{65}$ and Ge$_{25}$Sb$_{10}$Se$_{65}$ Amorphous
Thin Films: Sputtering and Pulsed Laser
Deposition for Near- and Mid-IR Applications,"
Int. J. Appl. Ceram. Technol. 8, 990–1000 (2011).
%\item M. Olivier, J.C. Tchahame, P. Němec, M. Chauvet, V. Besse, C. Cassagne, G. Boudebs, G.
\item M. Olivier, J.C. Tchahame, P. Nemec, M. Chauvet, V. Besse, C. Cassagne, G. Boudebs, G.
Renversez, R. Boidin, E. Baudet, and V. Nazabal,
“Structure, nonlinear, properties,
andphotosensitivity of (GeSe$_2$)$_{100-x}$(Sb$_2$Se$_3$)x
glasses”, Opt. Mat. Expr., 4, 525-540 (2014).
\item M. M. R. Elsawy and G. Renversez, “Study of
plasmonic slot waveguides with a nonlinear
metamaterial core: semi-analytical and numerical
methods”, Journal of Optics, 19, 075001 (2017).
\item M. M. R. Elsawy and G. Renversez, "Exact
calculation of the nonlinear characteristics of 2D
isotropic and anisotropic waveguides," Opt. Lett.
43, 2446-2449 (2018).
\item M. M. R. Elsawy and G. Renversez, “Improved
nonlinear slot waveguides using dielectric buffer
layers: properties of TM waves”, Opt. Lett., 41,
pp. 1542-1545 (2016).
\item S. Trillo and W. Torruellas (eds), Spatial Solitons,
SSOS, volume 82, Springer 2001.
\item R. Malendevich, L. Jankovic, G. Stegeman, and J.
S. Aitchison, "Spatial modulation instability in a
Kerr slab waveguide," Opt. Lett. 26, 1879–1881
(2001).
\item Y. Y. Lin, R. K. Lee, and Y. S. Kivshar,
"Transverse instability of transverse-magnetic
solitons and nonlinear surface plasmons," Opt.
Lett. 34, 2982–2984 (2009).
\item A. B. Aceves, J. V. Moloney, and A. C. Newell.
“Theory of light-beam propagation at nonlinear
interfaces. II. Multiple-particle and multiple-
interface extensions”, Phys. Rev. A 39, 1828
(1989).
\item J. Sánchez-Curto, P. Chamorro-Posada, and G. S.
McDonald, “Bright and black soliton splitting at
nonlinear interfaces”, Phys. Rev. A 85, 013836
(2012).
\item C. Cambournac, H. Maillotte, E. Lantz, J. M.
Dudley, M. Chauvet, “Spatiotemporal behavior of
periodic arrays of spatial solitons in a planar
waveguide with relaxing Kerr nonlinearity”, J.
Opt. Soc. Am. B, 19, 574-585 (2002).
\item G. Boudebs, S. Cherukulappurath, H. Leblond, J.
Troles, F. Smektala, F. Sanchez, “Experimental
and theoretical study of higher-order
nonlinearities in chalcogenide glasses”. Opt.
Comm., 219, 427-433 (2003).
\item R. W. Boyd, Z. Shi, I. De Leon, “The third-order
nonlinear optical susceptibility of gold”, Opt.
Comm., 326, 74-79 (2014).
\item H. Qian, Y. Xiao, Z. Liu, “Giant Kerr response of
ultrathin gold films from quantum size effect”,
Nat. Commun. 7, 13153 (2016).
\item A. Tuniz, S. Weidlich, M. A. Schmidt,
“Effectively Single-Mode Self-Recovering
Ultrafast Nonlinear Nanowire Surface Plasmons”,
Phys. Rev. Appl. 9, 044012 (2018).
\end{enumerate}
} %% fin du else du test 
%%%%%%%%%%%%%%%%%%%%%%%%%%%%%%%%%%%%%%%%%%%%%%%%%%%%%%%%%%%%%
\ifthenelse{\boolean{supinfo}}{%
%\onecolumn
\newpage
\appendix
\noindent \textbf{\Large Supplementary information of the article\\"Experimental demonstration of plasmon-soliton waves"}  
\tableofcontents
\newpage

\section{Details of the nonlinear mode profiles}
\label{sec:nonlinear-modes-FEM}
We can compute the nonlinear vector modes that propagate in a waveguide or photonic structure with a Kerr nonlinearity region as a function of the mode power using the model we have already developped using the finite element method (FEM) and the fixed power algorithm~\cite{Elsawy18nl-charact} (see the Methods section in the main article) based on Maxwell's equations . In this section we focus on two issues: the impact of the thickness $d$ of the silica buffer layer between the gold layer and the chalcogenide nonlinear layer on the type of modes, and the impact of the power carried by the nonlinear solution on the modes distribution.

\subsection{Impact of the  thickness $d$ of the silica buffer on the nonlinear mode type}
As already stated (see main article and quoted references), the thickness $d$ of the silica buffer plays a key role in the mode type independently of the considered linear or nonlinear regimes. In Fig.~2 of the main article, only profiles of the modulus squared of the electric field (intensity) along the X-axis for $y=0$ are provided for the main TE and TM modes for two  silica buffer thicknesses $d$.  In Fig.~\ref{fig:3D-profiles-TM-h10nm-h30nm}, a 3D view of the intensity profiles of the nonlinear modes with the same carried total power for  $d=10$ nm  and $d=30$ nm is given.

\begin{figure}[!bt]
  \centering
  \includegraphics[width=0.8\textwidth]{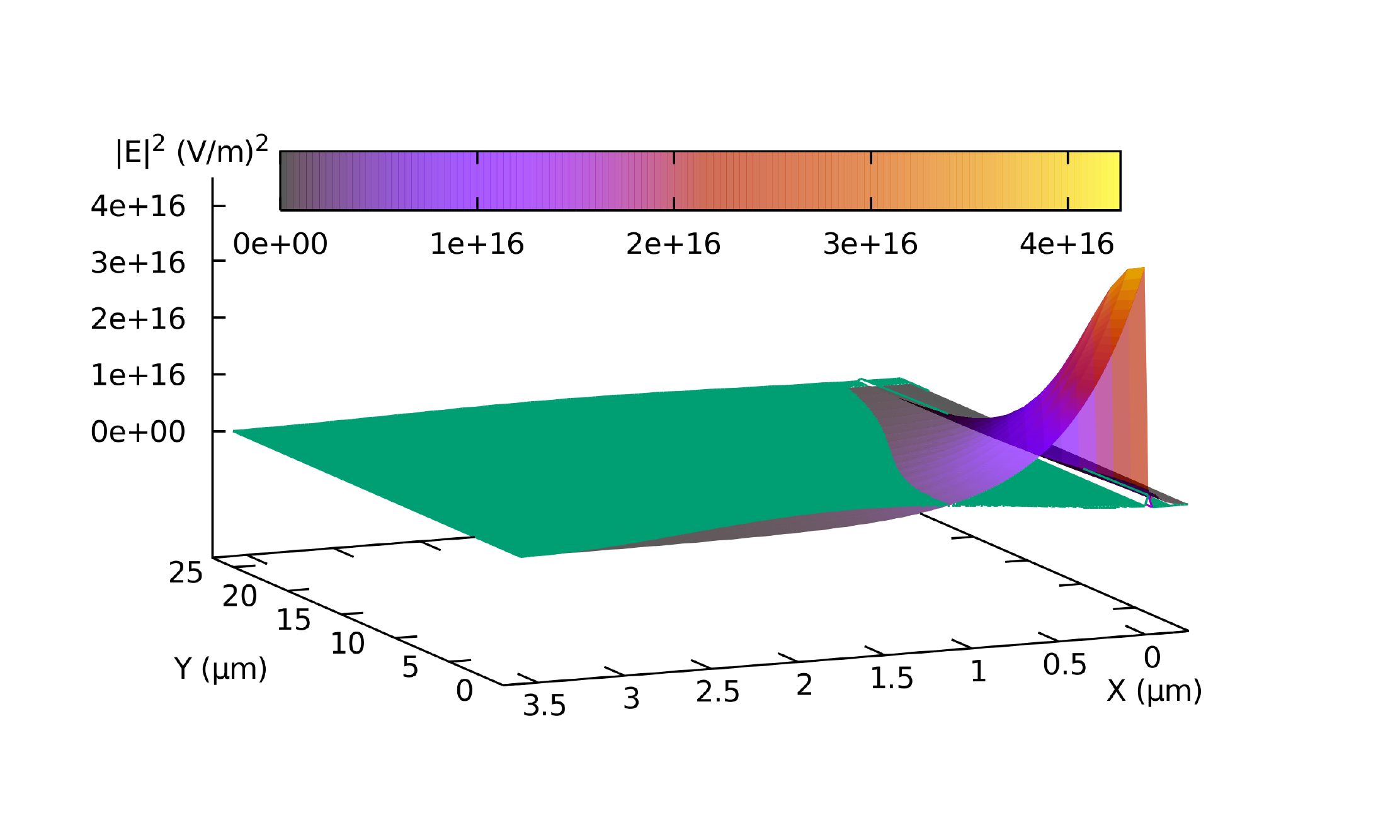}
  \caption{Intensity distribution of the nonlinear TM modes (only one half shown thanks to the symmetry axis along $y=0$) for $d= 10$ nm (with color scale) and for $d=$ 30 nm (green) with the same total power $P_{tot}=513 $ W.  
  The vertical scale is identical for the two modes. Note that for $d$=10 nm, the distribution in the silica buffer layer is not shown due to its large value as shown in the profile provided in Fig.~2 of the main article. See Fig.~\ref{fig:3D-profiles-TM-h10nm-only} for the view of the distribution in the full structure.} 
  \label{fig:3D-profiles-TM-h10nm-h30nm}
\end{figure}
\begin{figure}[!bt]
  \centering
  \includegraphics[width=0.8\textwidth]{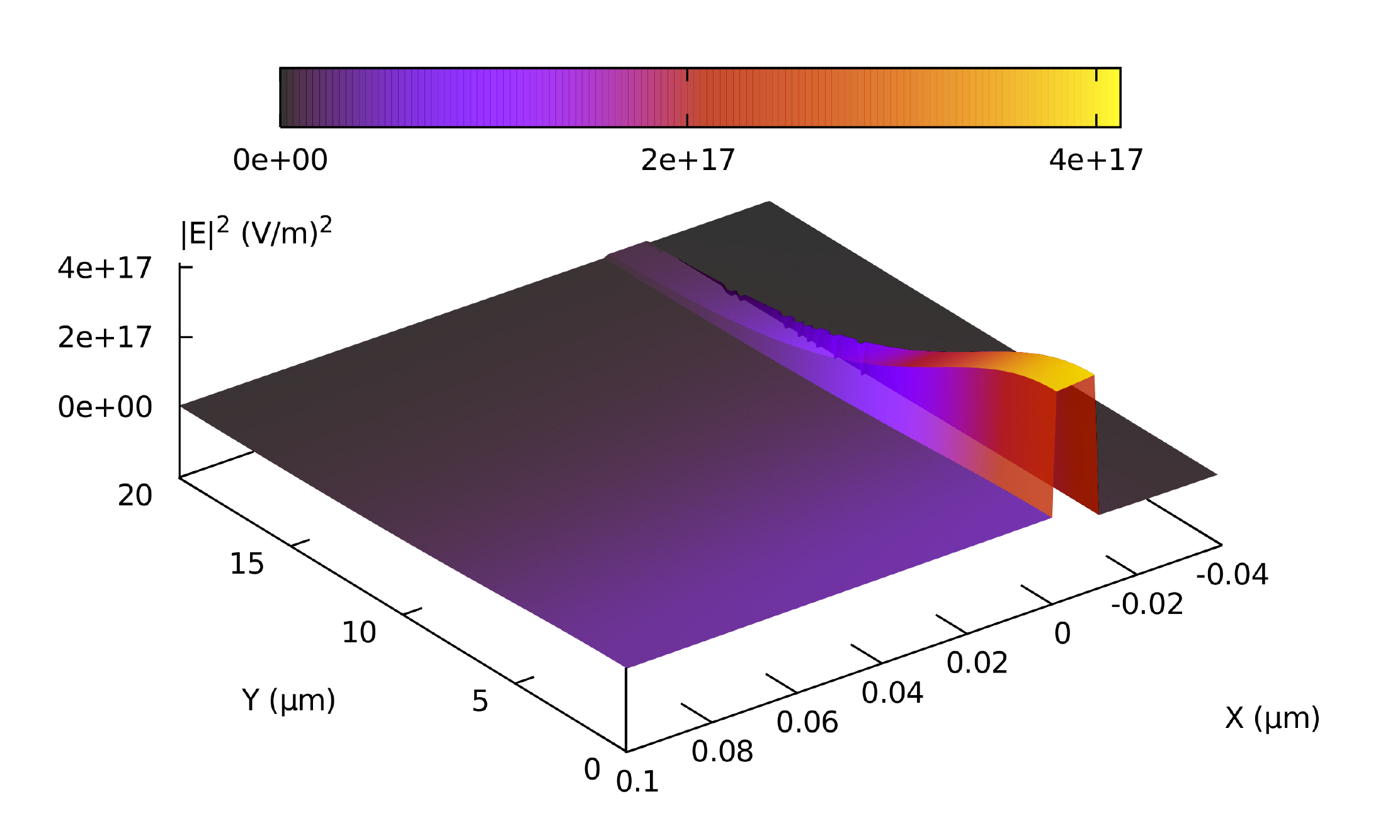}
  \caption{Intensity distribution of the nonlinear TM modes propagating inside the full PS for d=10 nm with a total power $P_{tot}=513 $ W (only one half shown thanks to the symmetry axis along $y=0$). Note that due to the large values of the intensity in the silica buffer layer, the intensity in the other regions are not well resolved (see Fig.~\ref{fig:3D-profiles-TM-h10nm-h30nm} for details of these regions.)}
  \label{fig:3D-profiles-TM-h10nm-only}
\end{figure}

First,  Figure~\ref{fig:3D-profiles-TM-h10nm-h30nm}  shows the 2D localization of the investigated modes induced by the nonlinearity especially along the Y axis despite the absence of any permittivity contrast in this direction in the initial structure.
Second, this figure also clearly illustrates the stronger localization of the electric field of the TM mode for a PS with a silica buffer layer of $d=10$ nm thickness (see Fig.~\ref{fig:3D-profiles-TM-h10nm-only} for a full view of this TM mode) compare to a PS with $d=30$ nm. 
It is this stronger spatial localization for both along the X-axis and the Y-axis that makes the improvement of the effective nonlinearity for the TM mode in presence of the PS compare to the ones of the three other cases studied experimentally in the main article: TM mode without PS, TE mode with and without PS.

For completeness, in Fig.~\ref{fig:3D-profiles-TM-TE-h30nm}, we provide a full 3D view of the intensity profiles of the TM and TE nonlinear modes for the same total power of $P_{tot}=513 $ W for $d=30$ nm. One can see that both polarizations have similar features in term of localization as stated in the main article.

\begin{figure}[!bt]
  \centering
  \includegraphics[width=0.7\textwidth]{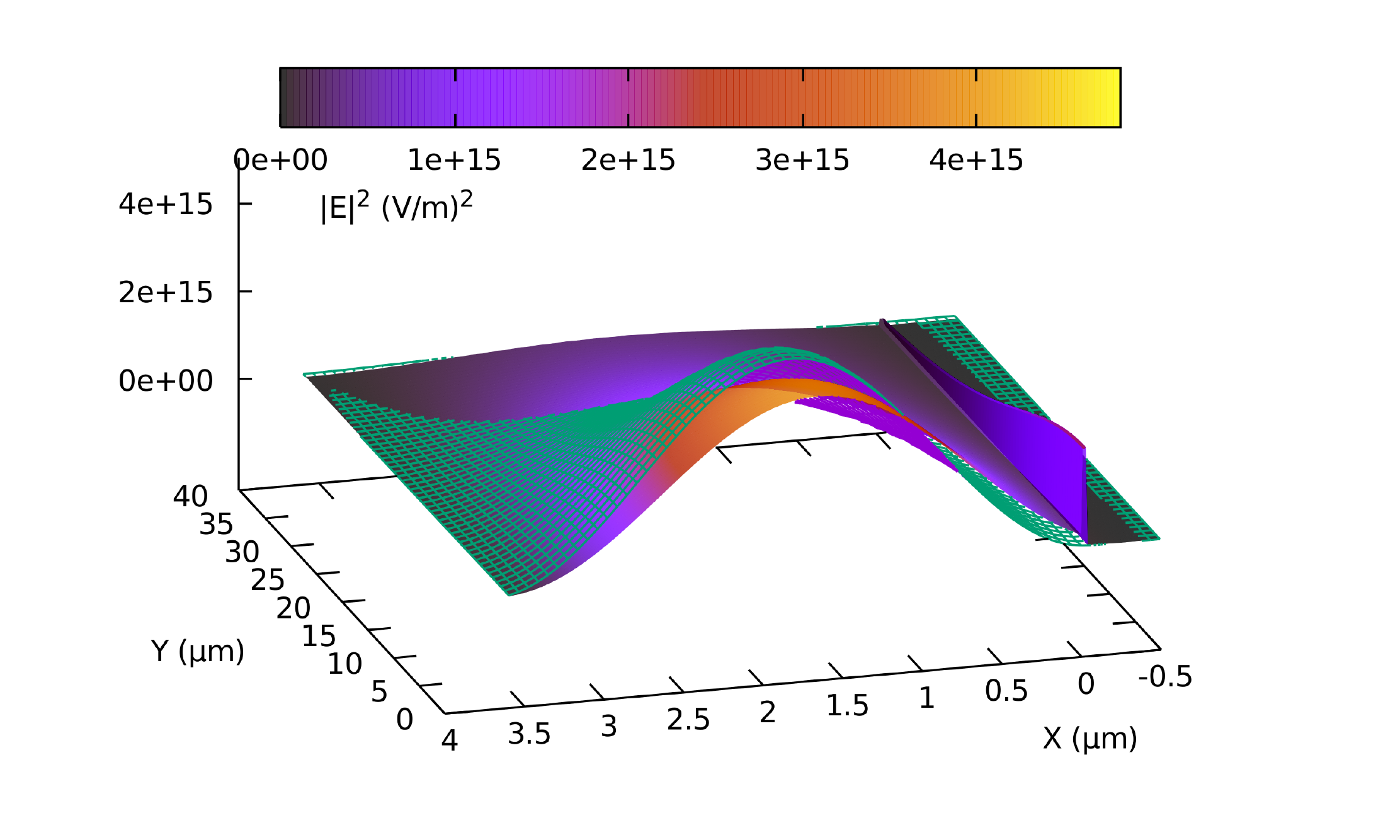}
  \includegraphics[width=0.7\textwidth]{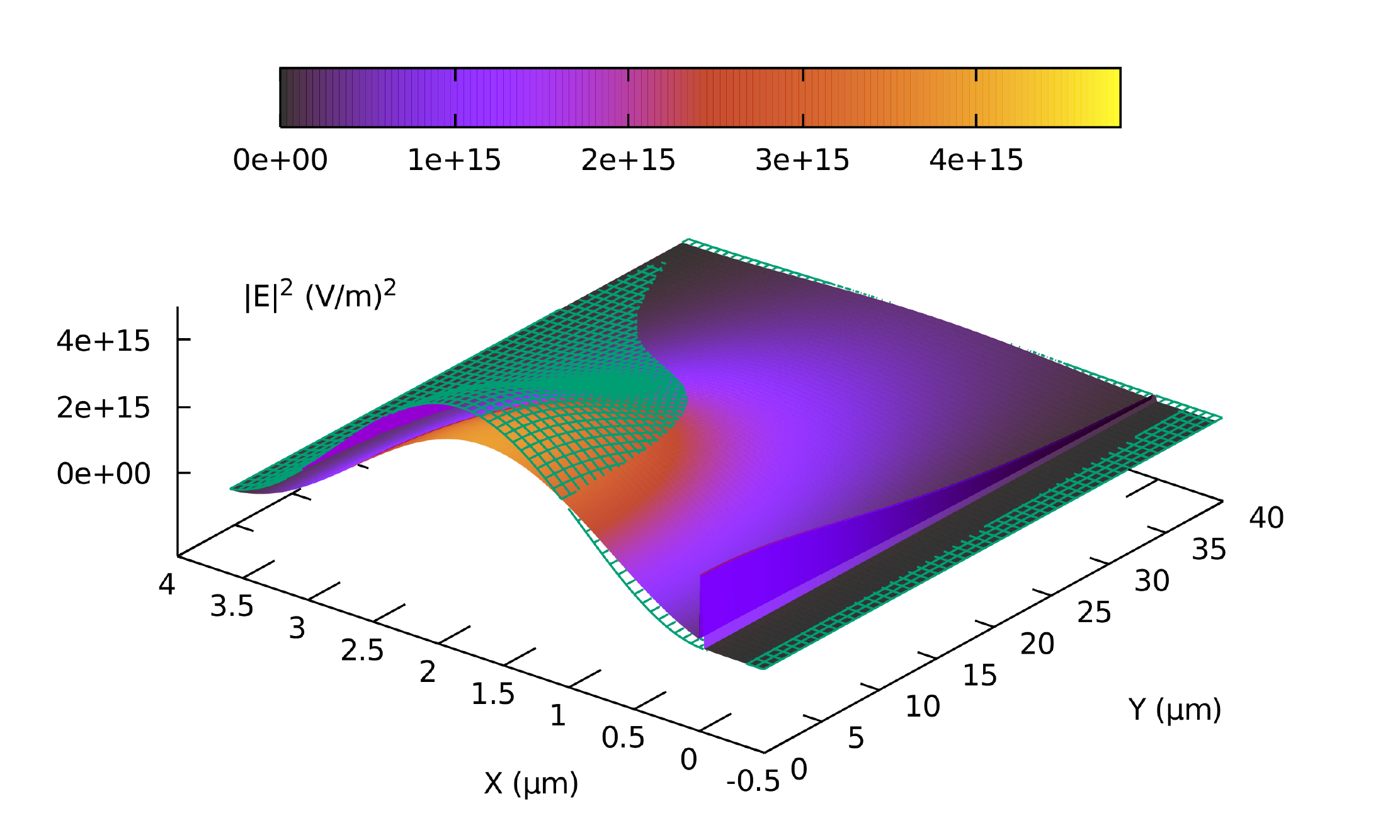}
  \caption{Intensity profiles of the nonlinear TM and TE modes for $d=30$ nm and a power $P_{tot}=513 $ W seen from two view-points. The TM mode is shown with the color scale while the TE mode is shown using the green mesh. The vertical scale is the same for the two modes.(only one half shown thanks to the symmetry axis along $y=0$)}
  \label{fig:3D-profiles-TM-TE-h30nm}
\end{figure}

\subsection{Impact of the total power on the nonlinear modes}
In nonlinear regime (at high intensity), the shape of the mode field profiles depends on the carried power.
Two normalized intensity profiles  of the nonlinear TM modes computed for  $P_{tot} $= 513 W and $P_{tot} $= 1108 W  are shown  in Fig.~\ref{fig:3D-profiles-TM-h10-2powers}. 

Along the X-axis, the mode width dependence on power is weak since it is mainly dictated by the layers layout (see Table~\ref{tab:eta} in the next section for a quantitative impact of this power dependence). The impact of power on the Y-axis FWHM is stronger but there is no need to insert this width in the propagation equation~\eqref{eq:improved-SNLE} since this equation describes by itself the focusing effect along the Y-axis due to the nonlinearity. Note that these FEM results cannot be compared directly with the experimental ones. Indeed, experimentally the propagation also includes the progressive beam reshaping toward a soliton profile since the input beam is fairly focused on the input facet of the structure and does not correspond to the self-coherent solution corresponding to the computed nonlinear modes.

\begin{figure}[!bt]
  \centering
  \includegraphics[width=0.75\textwidth,draft=false]{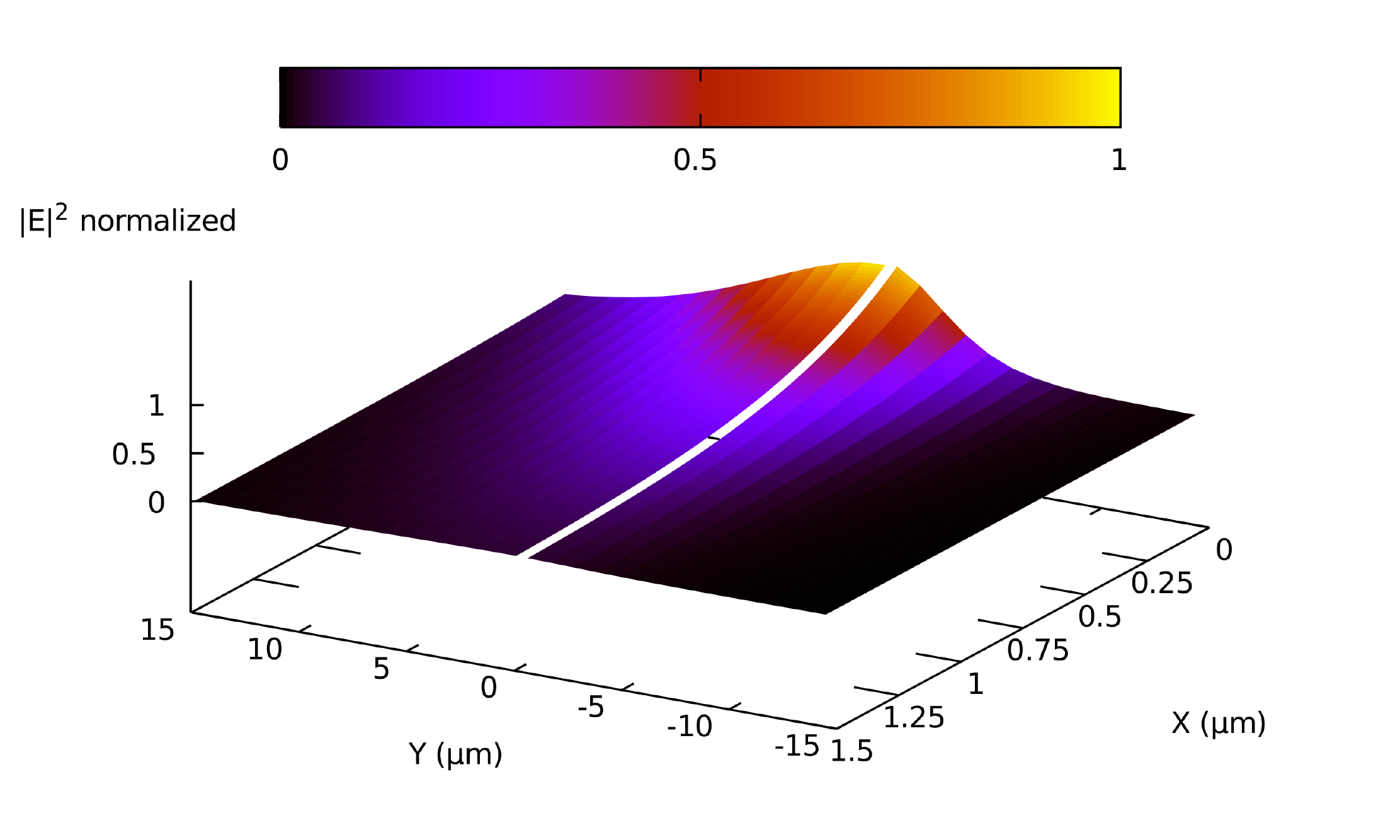}
  \caption{Normalized intensity distribution of the nonlinear TM modes for two different powers $P_{tot} $ for $d=10$ nm: $P=513$ W on the left part (positive Y), $P=1026$ W on the right (negative Y).(only one half shown thanks to the symmetry axis along $y=0$)}   \label{fig:3D-profiles-TM-h10-2powers}
 \end{figure}

\section{Details of the solved nonlinear propagation equation}
\label{sec:nonlinear-prop-equation-derivation}

\subsection{Model and equations}
\label{sec:nonlinear-prop-equation-derivation-model}
With our propagation simulations using a SNLSE (spatial nonlinear Schrödinger equation), we want to predict the spatial profile along Y-axis versus propagation along the Z-axis taking into account the spatial profile along the X-axis of the nonlinear mode in the studied region of the structure. To realize this task, we consider that the structure is step-wise constant along the Z-axis.
 In each discretized segment of the propagation, we determine the main mode using our FEM mode solver (see Methods. Simulations. Modal Analysis). The mode XY-distribution is determined together with its propagation constant in each segment. 

It is worth to remind that usually in basic spatial solitons studies~\cite{kivshar03,powers2012} only one transverse dimension is considered. In the present study since we need to take into account the spatial field profile along the X-axis that differs for  the dielectric type mode profiles and for the plasmonic type mode profile obtained only for the TM polarization with PS, we start from the nonlinear evolution equation where both $x$ and $y$ variables are considered. 
To get the final SNLSE involving only the transverse coordinate $y$ and the propagation direction coordinate $z$, we proceed in a similar way as it is done when the NLSE is derived in the temporal domain using a spatial integration in the cross section of the waveguide~\cite{agrawal89-nonlin-fiber-optics}. This study differs since we limit the spatial integration to the nonlinear region of the X-axis to keep the dependency along the Y-axis. We also renormalize the investigated amplitude denoted by $\tilde{A}(y,z)$ in order to get directly the optical intensity $I$ when one takes its modulus squared.
\begin{eqnarray}
  \label{eq:definitions}
  \mathbf{E}(\mathbf{r},t) &=& \frac{1}{2} \mathbf{\hat{u}} \left( F(x) A(y,z) \exp(-\imath (\omega_0 t - \beta_0 z))  + \mathrm{cc} \right) \\
  | \tilde{A}(y,z)|^2 &=&   |A(y,z)|^2  \frac{2}{ \varepsilon_{0} n c} 
\end{eqnarray} 
where $\mathrm{cc}$ means the complex conjugate and $\mathbf{\hat{u}}$ is a fixed unitary vector in the waveguide cross-section depending on the studied polarization.  $\tilde{A}(y,z)$ obeys the following SNLSE:
\begin{eqnarray}
    \DP{\tilde{A}(y,z)}{z} &=& \frac{\imath}{2 \beta_0} \DPn{2}{\tilde{A}(y,z)}{y}  - \frac{\alpha}{2} \tilde{A}(y,z) - \frac{\alpha_2}{2} |\tilde{A}(y,z)|^2 \tilde{A}(y,z) +  \frac{\imath \omega_0 n_2}{c} \eta^{1D}_{x} |\tilde{A}(y,z)|^2   \tilde{A}(y,z) \label{eq:improved-SNLE}\\
  \mbox{\textrm{with }} &  &  \nonumber \\
    \eta^{1D}_{x}   &\equiv& \int_{\mathrm{NL} \, \mathrm{layer}} F(x)^4 dx /\int_{\mathrm{NL} \, \mathrm{layer}} F(x)^2 dx 
  \end{eqnarray}
where $F(x)$ is the normalized spatial profile along $x$ of the mode of interest in the considered segment of the structure,  $\alpha$ is the linear loss coefficient, $\alpha_2$ the two-photon absorption nonlinear coefficient, and $n_2$ the Kerr nonlinear coefficient of the chalcogenide layer. The one-dimensional integrals defining  $  \eta^{1D}_{x}$ are computed only in the  nonlinear region.
These integrals are  computed using our FEM mode solver as the propagation constant $\beta_0$ and the full field profiles.

We call this  $\eta^{1D}_{x}$ parameter the opto-geometrical nonlinear factor.
Before pursuing the description of the $\eta^{1D}_{x}$ quantity, it is needed to explain how the experimental value of the material nonlinear coefficient $n_2$ has been obtained. In several cases, its value is derived from nonlinear propagation simulations using the two-dimensional spatial nonlinear Shrödinger equation where the transverse FWHM is computed and then compared to the measured ones as we have already done in reference~\cite{kuriakose17Measurement-ultrafast-optical-Kerr-effect-chalco} but the used equation did not take into account the field profile along the X-axis (axis perpendicular to the layer) and consequently the factor $\eta^{1D}_{x}$  was missing.  The used equation was actually the equation describing the spatial variation  of an optical beam  during the nonlinear propagation only in one transverse direction but not within a 2D cross section structure.  If we compare these two equations (Eq.~(1)  in reference~\cite{kuriakose17Measurement-ultrafast-optical-Kerr-effect-chalco}, or equivalent equation given in~\cite{powers2012} with Eq.~\eqref{eq:improved-SNLE} given in this Section~\ref{sec:nonlinear-prop-equation-derivation}), one can see that the $\eta^{1D}_{x}$ factor describing the field profile in the direction perpendicular to the propagation axis (Z) and to the axis associated with the second spatial coordinate of the SNLE (Y) appears only in the last term of Eq.~\eqref{eq:improved-SNLE}. 
This factor  $\eta^{1D}_{x}$ is similar to the   $\eta$ one described in the derivation of the  nonlinear Schrödinger equation in the temporal domain for optical fibers by formula (5.26) in Okamoto's book~\cite{OKAMOTO2006-chapter5}. But in this later case, the spatial integration is realized on the full 2D cross section (not only on the nonlinear fiber core) in order to approximate its value by a factor 1/2 for the HE$_{11}$ mode of a circular step-index fiber under the normal operating condition of such fiber. This is not the case in our study where we integrate only along the dimension (X-axis) corresponding to the thickness of the nonlinear layer.
The values of  $\eta^{1D}_{x}$  for the studied structures, for the main TE and TM modes with  and without the PS are given in Table~\ref{tab:eta}.  These values are computed using the nonlinear fixed power algorithm described in reference~\cite{Elsawy18nl-charact}, see also Section~\ref{sec:nonlinear-modes-FEM} of this Supplementary information and the  Methods Section (Simulations. Modal Analysis) of the main article.

\begin{table}
  \centering
  \begin{tabular}[t]{|c|c|c|c|}
    \hline
     $P_{tot}$ (W) & 513 & 1108 & 2216 \\   \hline
    TM with PS & 1.23&  1.26 & 1.37\\ \hline
    TE with PS & 0.40 & 0.40  &  \\ \hline
    TM without PS & 0.39 & 0.39 & 0.39 \\ \hline 
    TE without PS & 0.40 & 0.40 & \\  \hline
    {\small ratio: TM with PS/ } & 3.15 & 3.23  & 3.51 \\  
      TM without PS & & & \\ \hline
  \end{tabular}
  \caption{Opto-geometrical nonlinear factor $\eta^{1D}_{x}$, for the main TE and TM modes with and without the PS for several values of the mode total power and a silica buffer layer thickness $h=10$ nm.}
  \label{tab:eta}
\end{table}
The TE mode without PS and TM mode without PS correspond to the cases used to evaluate the nonlinear coefficient $n_2$ in reference~\cite{kuriakose17Measurement-ultrafast-optical-Kerr-effect-chalco}. As a result, the previously provided $n_2$ takes into account the mode related quantity   $\eta^{1D}_{x}$ for the TE and TM cases without PS, its value is numerically evaluated to approximately 0.4. For the TM mode with PS, $\eta^{1D}_{x}$ is between 3 and 3.4 times higher than for the other cases. It slightly increases with the mode power as shown in  Table~\ref{tab:eta}. 
Consequently, the effective nonlinearity improvement for the TM mode with PS is at least three fold the ones without plasmonic enhancement. 
We found that in order to quantitatively reproduce the different experimental results for the three experimentally tested configurations (I, II, and III) we have to use an opto-geometrical nonlinear factor not  three times higher, as evaluated by the FEM simulations, but eight times higher. We mainly attribute this discrepancy to our modelling approach (see Section~\ref{sec:discussion-nonlinear-propa-model} in this document and the last paragraph of the section Simulation-Discussion of the main article). In other words, our modelling approach underestimates the observed nonlinearity in the TM case with PS. Among possible explanations, we didn't take into account the gold nonlinearity. 
%and the Kerr nonlinearity may not be not treated as an instantaneous for 200 fs pulses. Cet argument n'est pas évident car nous devrions donc également le voir dans les simuls sans PS ?
 It is worth pointing out that in the linear regime our simulations recover the expected results from linear Gaussian beam propagation. 

\subsection{Numerical implementation}
 
We solve Eq.~\eqref{eq:improved-SNLE} using the Runge Kutta method of order 4 with the Interaction Picture (RK4IP)% interaction ou intraction + il manque pas une ref. ici ? pb réglé, la ref est plus loin Hult07
 method, initially designed  to solve the NLSE in the temporal domain, but adapted in the present case to the spatial domain since the RK4IP method has been proven to be more accurate and more efficient than the split-step Fourier method~\cite{Hult07}. This task is relatively straightforward since there is no spatial dispersion to take into account due to the fact that we investigate only the beam propagation at a single wavelength (1.55 $\mu$m) while to solve the time equation one must take into account the frequency dependency of material properties to get accurate results. 
In  addition to improve the efficiency of our numerical implementation, we have used the local error adaptative step size method that allows to adapt continuously the spatial step size in the propagation direction but keeping the local error below a fixed threshold~\cite{Sinkin03}. We have already used it successfully to study supercontinuum generation in highly nonlinear chalcogenide microstructured optical fibers~\cite{elamraoui10spectral-broadening-suspended-core-MOF}.

\section{Impact of parameter variability for the simulations of configuration I}
\label{sec:impact-parameters}
In this section, we describe the impact of parameters values on the simulations results describing configuration I.
\subsection{Influence of silica buffer layer thickness $d$}
\label{sec:impact-parameters-d}

We describe the influence of the silica buffer thickness when it is below the critical size ensuring a plasmonic behaviour for the TM mode with the PS. The targeted thickness $d$ of 10 nm was obtained in the fabricated sample (see Fig.~\ref{fig:geometry-general-scheme}c in the main article). Nevertheless, it is possible that this thickness $d$ is not constant along the fabricated PS. For instance, if $d=8$ nm instead of 10 nm, the corresponding mode  for $P$=513 W changes compared to the one computed for  $d=10$ nm: the $\eta^{1D}_{x}$ is now equal to  1.56 instead of 1.23. This implies that the ratio $\eta^{1D}_{x}$(TM with PS)$/\eta^{1D}_{x}$(TM without PS) increases from 3.2 to 4. Consequently, the effective nonlinearity is increased when $d$ decreases from 10 nm to 8 nm.

\subsection{Influence of $n_2$ and of the ratio $\eta^{1D}_{x}$(TM with PS)$/\eta^{1D}_{x}$(TM without PS)}
\label{sec:impact-parameters-d}

 In Fig.~\ref{fig:numerical-simulations-configI-impact-parameters}, the FWHM is computed from nonlinear propagation simulations for configuration I as a function of the input beam intensity for several sets of  $n_2$ and ratio $\eta^{1D}_{x}$(TM with PS)$/\eta^{1D}_{x}$(TM without PS) values including the one given in Fig.~\ref{fig:numerical-simulations}b of the article (purple curves shown in Fig.~\ref{fig:numerical-simulations-configI-impact-parameters}). As stated in ref.~[22], for a GeSbSe chalcogenide glass of similar composition and for an identical laser source as in the present study, the estimated uncertainty on $n_2$ is 28.5\%. This statement induces that the $n_2$ value  is typically  in the interval $[4-7]  \, 10^{-18}$ m$^2$/W. 
 % l'interval donné ici est plus large que dans les simuls ?
 One can see that the simulation results differ only quantitatively and that the FWHM variations stay in the correct range. To simplify the discussion in the article, we use  the $n_2$ value provided by the corrected results from reference~[22] \textit{i.e.}  $5.5 \, 10^{-18}$ m$^2$/W but one can see that if we choose  $4.5 \; 10^{-18}$ m$^2$/W, we obtain simulation results (green curves in Fig.~\ref{fig:numerical-simulations-configI-impact-parameters}) nearest to the experimental results (black curves in Fig.~\ref{fig:numerical-simulations-configI-impact-parameters}).

\begin{figure*}[!bt]
\centering
\ifthenelse{\boolean{figures}}{%
\includegraphics[width=0.9\columnwidth,clip=true,trim= 0 0 0 0]{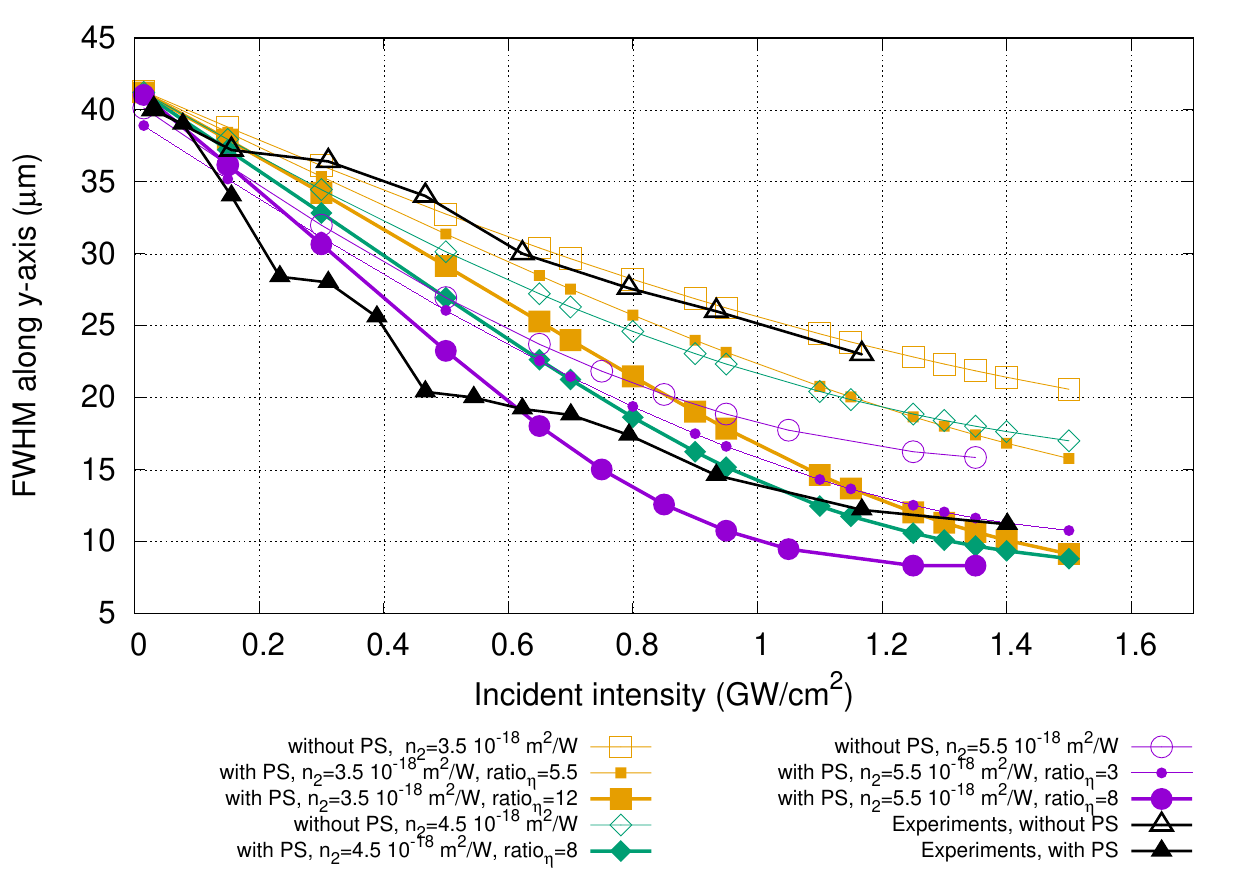}
}{}
\caption{Comparison of the computed FWHM for configuration I as a function of the input beam intensity for several sets of  $n_2$ and $\eta^{1D}_{x}$(TM with PS)$/\eta^{1D}_{x}$(TM without PS) ratio values, the experimental results are also shown in black for the TM case with (filled triangles) and without PS (empty triangles).}
\label{fig:numerical-simulations-configI-impact-parameters}
\end{figure*}

\section{Complementary results for configuration II}
\label{sec:nonlinear-prop-at-low-power}

\subsection{Propagation simulations for configuration II at low power}
In Fig.~5, we provide a color map of the beam intensity evolution along the Y-axis versus propagation distance inside the full structure for the TM polarization for configuration II with h=500 $\mu$m for an input average intensity of 1.25 GW/cm$^2$.

In order to complete these results and to validate the effect of the input intensity an additional color map is given in Fig.~\ref{fig:color-map-confi2-low-power} for same parameters except an input incident average intensity of 15 MW/cm$^2$. 
\begin{figure}[!bt]
  \centering
  \includegraphics[width=0.75\textwidth]{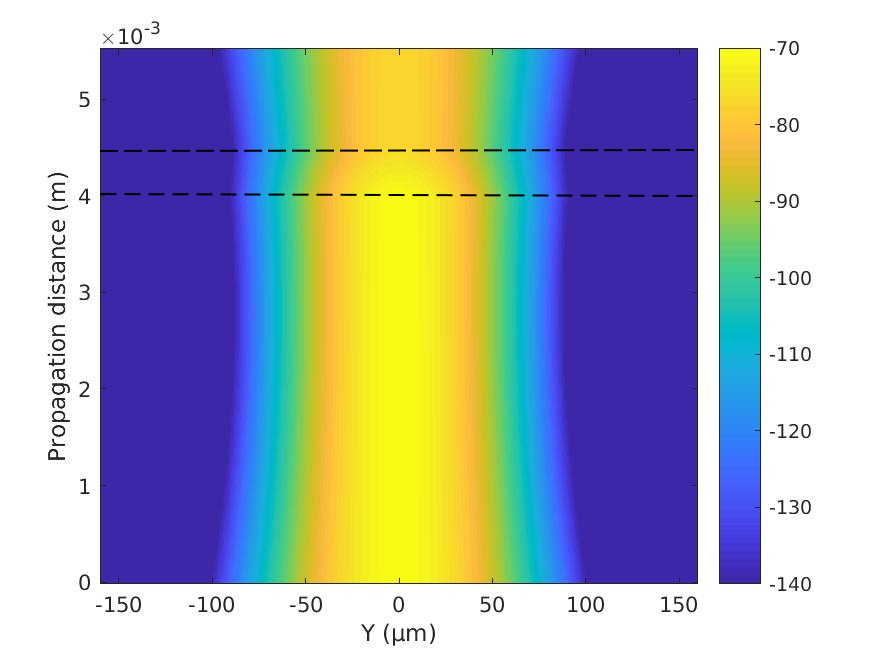} 
  \caption{Color map in log scale of the beam intensity profile along the Y-axis versus propagation coordinate inside the full structure for a TM polarization for configuration II with h=500 $\mu$m. Input intensity of 15 MW/cm$^2$. The dotted line represents the limits of the PS.} %
  \label{fig:color-map-confi2-low-power}
\end{figure}

In this figure, one can first see the focusing of the input beam on the initial section of the full structure (before the PS). This weak focusing is mainly due to the phase profile of the input beam~\cite{powers2012} since we must impose a spatial chirp to take into account the beam focusing induced by the microscope objective used to couple light in the structure.

Second, a focusing effect is visible under the PS structure region attributed to the enhanced  nonlinearity. Note that such a self-focusing takes place even for such a low intensity. Third, as the beam exits from the PS, a defocusing occurs due to the combination of the power loss induced by the higher propagation losses in the PS region and of the enhancement-free nonlinearity present away from the PS.

\subsection{Comparisons between experimental results and numerical simulations for configuration II as a function of the PS length}
\label{sec:comparisons-configII}
In Fig.~\ref{fig:config2-FWHM-versus-I-different-h}, we compare the experimental results and numerical simulations for configuration II as a function of the length $h$ of the PS. The FWHM of the output intensity profile in the Y-direction is depicted as a function of the incident average intensity for three different lengths $h$ of the PS (see Fig.~1a-b in the main article). 

One can see that, for the largest $h$ equal to $500 \; \mu$m, the computed Y-axis FWHM of the beam at the output facet of the full structure are in fair agreement with the experimental results. Nevertheless, it must be pointed out that the profiles along Y-axis are no longer simple Gaussian or hyperbolic secant at high intensities. At high power regime, the  profiles have multiple symmetric peaks as observed experimentally (Fig.~5 in the main article) and numerically (Fig.~6 in the main article). Consequently the FWHM is neither sufficient nor adequate to describe the features of the beam profile along the Y axis. The reported defocusing of the output beam observed experimentally for the highest intensity is also linked to the generation of these lateral peaks. This is seen above 1 GW/cm$^2$ for configuration II for $h=500 \; \mu$m both in the experiments and in the simulations. Note that this defocusing effect was not seen in the FWHM experimental results in configuration I. The main reason is that the PS is located at the imaged output face of the full structure and consequently no diffraction can occur contrary to configuration II where the beam propagates some distance in the dielectric structure before its analysis. Nevertheless, when we increase the power of the input beam in the numerical simulations above 2 GW/cm$^2$ in configuration I for $h=660 \; \mu$m, the generation of small lateral peaks are observed similarly as for configuration II. We infer that in the experiments for configuration I, the intensity reaching the PS structure is too low to induce the lateral peaks, and consequently the broadening regime of beam is not yet reached.

\begin{figure}[!bt]
  \centering
  {\includegraphics[width=0.9\textwidth]{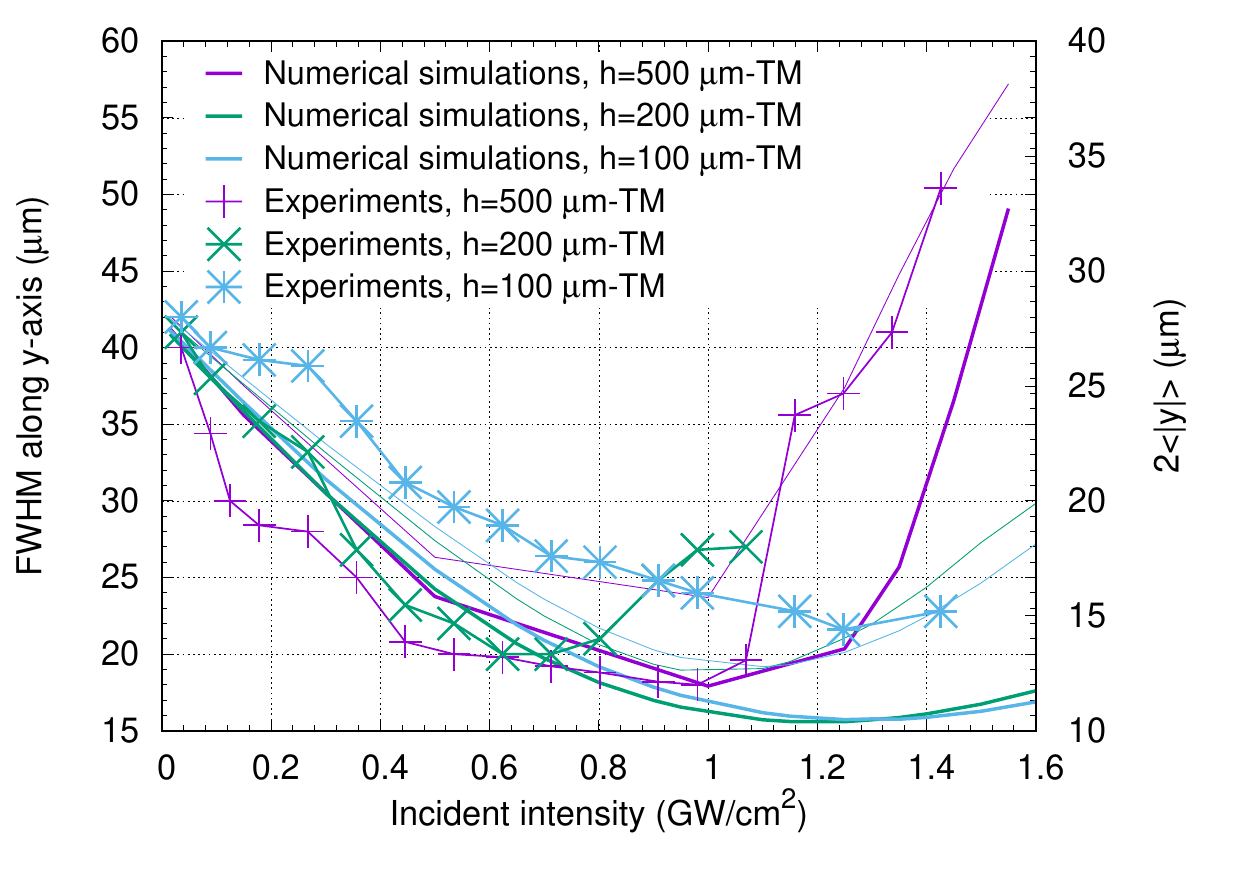}}
  \caption{Computed and measured output beams FWHM along Y-axis for configuration II as a function of the input intensity for three different length $h$ of the PS. On the right vertical scale, the corresponding computed mean full width $2 <|y|>$ of the output beam along the Y-axis are given  (see the text for its definition) using thin curves.}
  \label{fig:config2-FWHM-versus-I-different-h}
\end{figure}
For $h=200 \; \mu$m, one can see that the measured and simulated results are also in good agreement up to 0.8 GW/cm$^2$. Above this intensity, the simulations cannot reproduce the increase of the FWHM followed by a short plateau (only two experimental points for this plateau).  For $h=100 \; \mu$m, the experimental curve exhibits a gradual decrease followed by a final stabilization of the measured FWHM of the output beam. As already mentioned at the beginning of section~\ref{sec:comparisons-configII} and as illustrated in Fig.~5 of the article,  the beam profiles contain multiple lateral peaks when the incident intensity becomes large. When this regime starts the FWHM does not reflect it since the smaller lateral peaks do not modify the FWHM. Consequenlty, the discrepancies between the experimental results and the numerical simulations increase with the incident intensity, up to the intensity requested to generate the high lateral peaks.

To allow our simplified model to take into account at least qualitatively the impact of the multiple lateral peaks on the spatial spreading of the output beam widths along the $Y$-axis we compute the quantity $2 <|y|>$ defined by:
$$ 2 <|y|> \equiv 2 \int_{Y-section} |y|  |E|^2(y,z_{out}) dy / \int_{Y-section} |E|^2(y,z_{out}) dy $$
where $z_{out}$ is the $z$-coordinate of the output face of the full structure.
This quantity  $2<|y|>$ corresponds to the mean full width along the $Y$-axis of the symmetric profiles under investigation, it can capture the influence of the  multiple lateral peaks even if their maximum  values do not reach the half maximum of the central peak.  The corresponding curves for the three structures we studied in configuration II ($h=500, 200, 100 \, \mu$m) are the thin ones in Fig.~\ref{fig:config2-FWHM-versus-I-different-h} linked to the right vertical scale. As it can be seen, the results are  in better agreement qualitatively with the experimental data than the direct output of the simulated FWHM. They show that the increase of the width of the output beam appears for smaller values of the incident intensities compared the results given by the simulations of the FWHM even if quantitatively the increase observed for h=200 $\mu$m as a function of the incident intensity is not obtained from  0.7 GW/cm$^2$ but from 1.1 GW/cm$^2$.

\section{Complementary results for configurations I, II, and III}
\label{sec:nonlinear-prop-at-low-power}

In Fig.~\ref{fig:FWHM-versus-I-3configs}, we provide the comparisons of the computed and measured Y-axis FWHM output beams for the three configurations I, II, III for the largest $h$ value of each configuration as a function of the input beam intensity. As it can be seen, the computed FWHM curves follow partially the ones of the experimental data except for configuration III where the incident intensity of the PS segment is larger due to the shorter distance (1 mm) between the  input facet of the structure and the PS compared to  configuration II (4 mm) and configuration III (4.5 mm). For configuration III, the numerical results don't show any increase for the FWHM. Nevertheless, if we look at the results provided by mean full width  $2<|y|>$ of the output beam, we see that we obtain a better qualitative agreement for the comparisons, for configurations I and II the simulation results stay globally correct while, for configuration III, we get the expected increase of the output beam width even if it appears for larger incident intensities compared to the experimental data.
\begin{figure}[!bt]
  \centering
  {\includegraphics[width=0.9\textwidth]{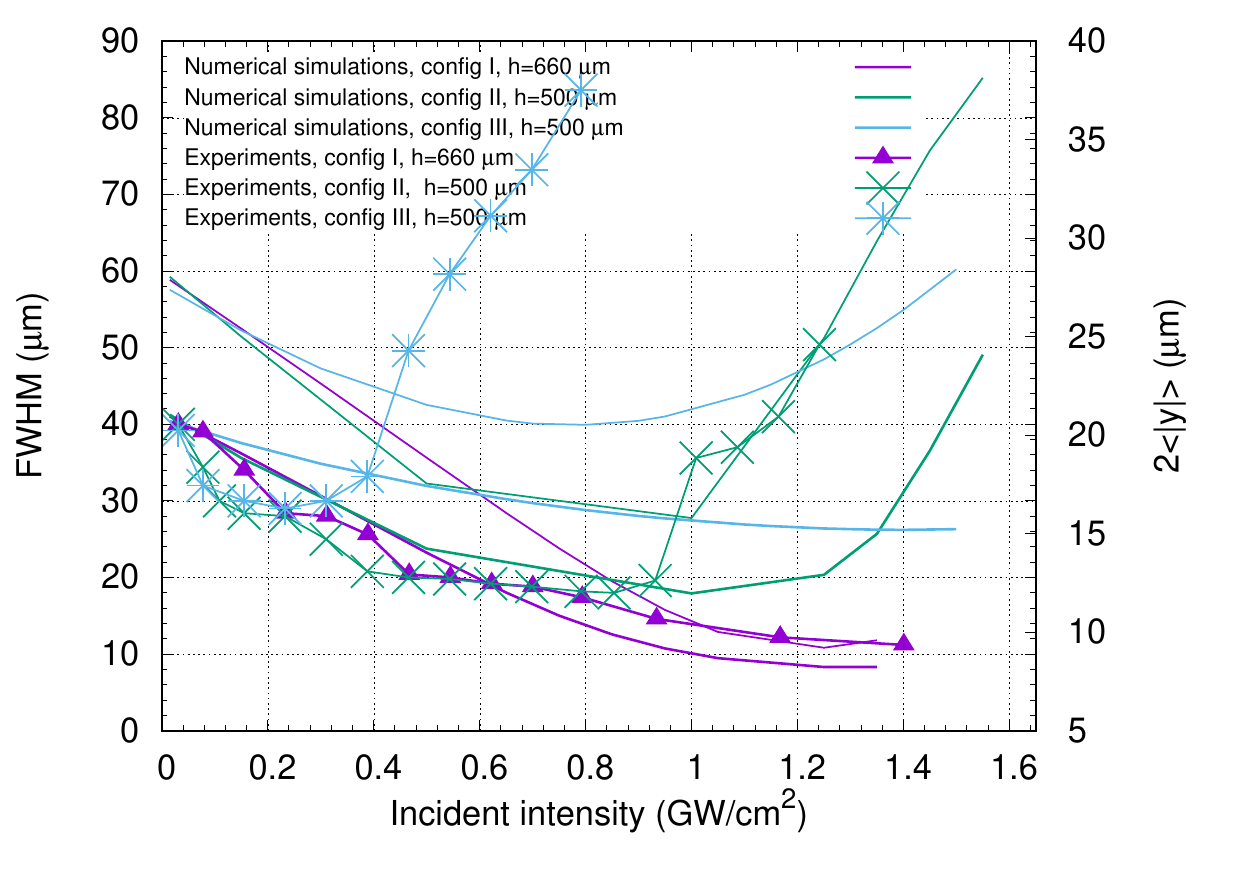}}
  \caption{Computed and measured Y-axis FWHM output beams for the three configurations I, II, III as a function of the input beam intensity. On the right vertical scale, the corresponding computed mean full width  $2 <|y|>$  of the output beam along the Y-axis are given  (see the text for its definition) using thin curves.}
  \label{fig:FWHM-versus-I-3configs}
\end{figure} %% 

\section{Discussion on the limitations of the SNLSE}
\label{sec:discussion-nonlinear-propa-model}
In this section, we extend the discussion summarized in the main article.
It must  be pointed out that the validity of our model does not extend to very high intensities. Indeed, in addition to the influence of higher order nonlinearities, the incertitudes on the considered nonlinear parameters ($n_2$, $\alpha_2$) may have a larger impact as one moves away from the linear regime. Furthermore, the hypothesis on the propagation model themselves are also becoming less and less valid as the incident intensity increases.  A full treatment of the Kerr nonlinearity considering the Maxwell's equations a~\cite{Akhmediev93,chen97maxwell-equations-vector-nonlinear-schrodinger-equation,drouart08beyond-townes-soliton} is more adequate than the Schrödinger equation. 
We must also remind that in the present form the solved SNLSE includes only one mode in the description of the nonlinear wave propagation as it is the case in the simple NLSE where the waveguide is usually assumed to be single-mode~\cite{agrawal89-nonlin-fiber-optics}. At high intensities this description may be no longer valid.

The limitations of the SNLSE can be at the origin of the observed discrepancies between the simulations results and the experimental ones. As an example, for configuration I and II, even if the numerical spatial profiles for the fields  contain symmetric multipeaks like  the recorded ones and have quite similar FWHM, they  differ quantitatively: their lateral peak positions being more distant to the beam center than the experimental ones for the same incident intensity. This phenomenon is similar to the predictions of 2D FDTD simulations of homogeneous Kerr media in which above some intensities the incident beam does not form a unique spatial soliton but a central peak with several lateral beams that diverge but with smaller angles than the ones seen in Fig.~\ref{fig:numerical-simulations}  (see appendix~C in~\cite{these-wiktor14}). Nonlinear FDTD simulations, eventually 3D ones, are needed to investigate more precisely these differences. Ultimately, the full numerical study of the plasmon-soliton waves should be realized within a nonlinear 3D approach. A supplementary reason not mentioned above comes from the layered geometry of the investigated structure. In our 2D SNLSE model the impact of the layered structure is only taken into account through the  opto-geometrical nonlinear factor $\eta^{1D}_{x}$, the weak variation of the propagation constants and the losses. The computational ressources to realize such 3D simulations is huge due to the large sizes of the studied structures. For a homogeneous cartesian grid of $\lambda/10$ sampling at $\lambda=1.55 \mu m$ and considering only half of the structure thanks to the symmetry properties, the order of magnitude of cells is 5 $\, 10^{6}$. This estimation is done without taken into account the grid size to correctly describe the nanoscale silica  buffer and gold layers (10 grid cells to described the silica layer thickness means a 1 nm grid size, to be compared to the 0.155 µm of the coarser grid). Consequently, such simulations cannot be realized efficiently realized without a non-uniform grid nonlinear 3D FDTD software. Furthermore, the above estimations do not consider the supplementary computational cost generated by the treatment of the nonlinearity in the FDTD code as it is explained in reference~\cite{taflove13}. 

\section{Test of thermal effects}
\label{sec:chopper-test}
In order to test possible thermal effects on the fosusing properties of the beam we have followed the procedure used in the litterature~\cite{Falconieri1999,Gnoli05}: to lower the time average power while maintaining the peak power an optical chopper was inserted on the beam path. In a previous work on spatial soliton in a chalcogenide waveguide of similar composition,we have already proven that the thermal effects is  negligible both at 1550 nm and at 1200 nm using the same femtosecond laser~\cite{kuriakose17Measurement-ultrafast-optical-Kerr-effect-chalco}.  
In order to check this behavior for the present chalcogenide structure, we repeated the chopper experiment with a decrease by 40\% of the time average intensity for the configuration III for $h$ = 500 $\mu$m. The results are given in Fig.~\ref{fig:chopper-impact}. Within the experimental errors, there is no change in the measured FWHM as a function of the incident intensity. These results confirm that thermal effect can be neglected in all the studied configurations since the PS of configuration III receives the highest power due to its nearest position from the input facet.  %% 

\begin{figure}[!bt]
  \centering

  \includegraphics[width=0.75\textwidth]{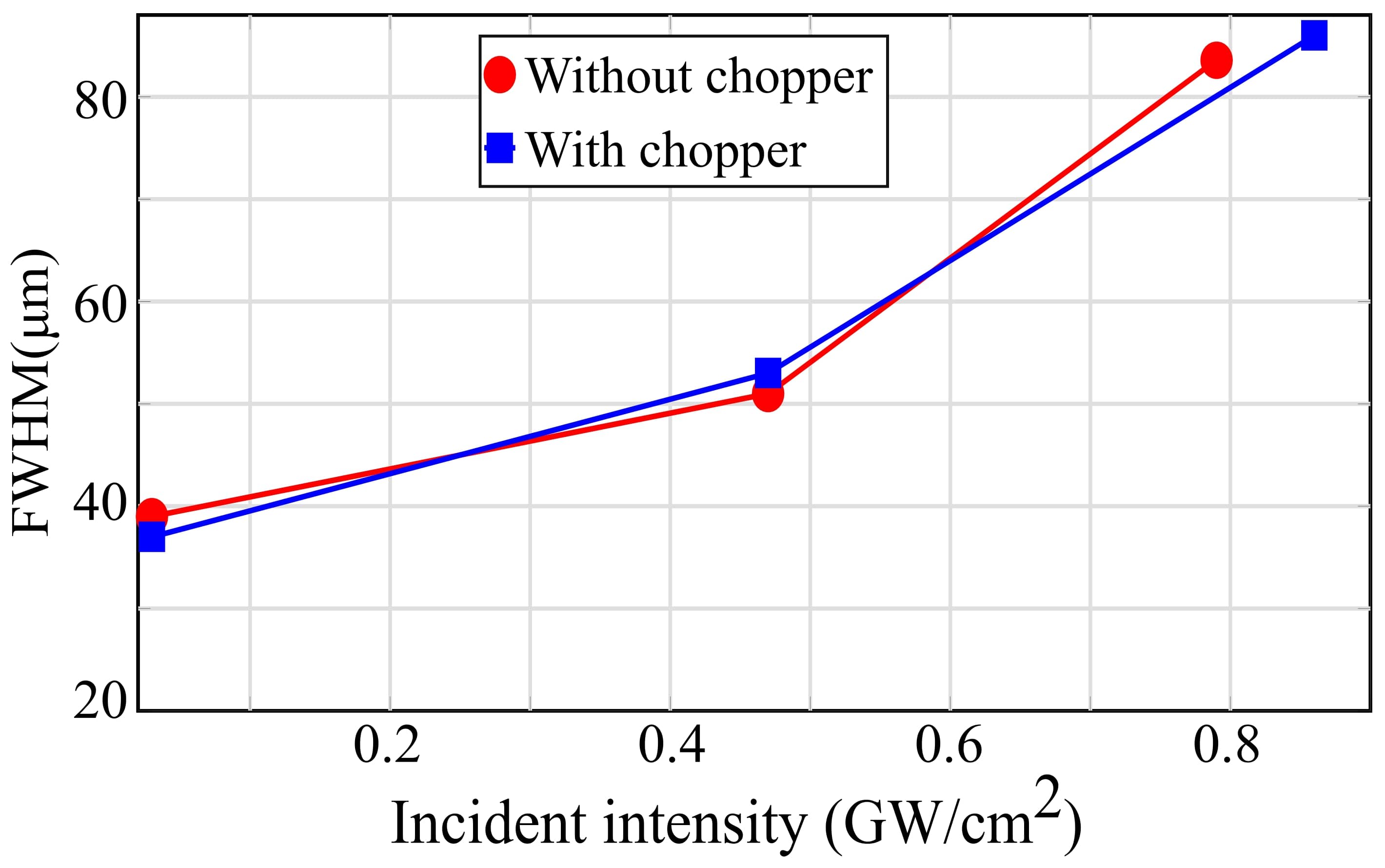}
  \caption{Evolution of the output beam FWHM as a function of incident intensity for configuration III with and without optical chopper.} %
     \label{fig:chopper-impact} 
\end{figure}

%%%%%%%%%%%%%%%%%%%%%%% References with titles
\newpage

%\vspace{1.0cm}
\addcontentsline{toc}{section}{References}%
\ifthenelse{\boolean{arxiv}}
{

%%  \bibliography{plasmon-soliton-paper-V9-with-supplement-info_biber}
}%else du test
{
  \printbibliography

\begin{thebibliography}{10}

\bibitem{Elsawy18nl-charact}
M.~M.~R. Elsawy and G.~Renversez.
\newblock Exact calculation of the nonlinear characteristics of {2D} isotropic
  and anisotropic waveguides.
\newblock {\em Opt. Lett.}, 43(11):2446--2449, 2018.

\bibitem{kivshar03}
Y.~S. Kivshar and G.~P. Agrawal.
\newblock {\em Optical Solitons, From Fibers to Photonic Crystals}.
\newblock Academic Press, 2003.

\bibitem{powers2012}
Peter~E. Powers and Joseph~W. Haus.
\newblock {\em Fundamentals of Nonlinear Optics}.
\newblock CRC Press, 2nd edition edition, 2017.

\bibitem{agrawal89-nonlin-fiber-optics}
G.~P. Agrawal.
\newblock {\em Nonlinear fiber optics}.
\newblock Academic Press, 3rd edition, 2001.

\bibitem{kuriakose17Measurement-ultrafast-optical-Kerr-effect-chalco}
T.~Kuriakose, E.~Baudet, T.~Halenkovič, M.~M.R. Elsawy, P.~Němec, V.~Nazabal,
  G.~Renversez, and M.~Chauvet.
\newblock Measurement of ultrafast optical kerr effect of ge-sb-se chalcogenide
  slab waveguides by the beam self-trapping technique.
\newblock {\em Opt. Comm.}, 403:352 -- 357, 2017.

\bibitem{OKAMOTO2006-chapter5}
Katsunari Okamoto.
\newblock Chapter 5 - nonlinear optical effects in optical fibers.
\newblock In Katsunari Okamoto, editor, {\em Fundamentals of Optical Waveguides
  (Second Edition)}, pages 209 -- 259. Academic Press, second edition edition,
  2006.

\bibitem{Hult07}
J.~Hult.
\newblock {A} {F}ourth-{O}rder {R}unge-{K}utta in the {I}nteraction {P}icture
  {M}ethod for {S}imulating {S}upercontinuum {G}eneration in {O}ptical {F}iber.
\newblock {\em IEEE J. Lightwave Technol.}, 25(12):3770--3775, 2007.

\bibitem{Sinkin03}
J.Zweck O.~V.~Sinkin, R.~Holzlohner and C.~R. Menyuk.
\newblock Optimization of the {S}plit-{S}tep {F}ourier {M}ethod in {M}odeling
  {O}ptical {F}iber {C}ommunications {S}ystems.
\newblock {\em IEEE J. Lightwave Technol.}, 21(1):61--68, 2003.

\bibitem{elamraoui10spectral-broadening-suspended-core-MOF}
M.~El-Amraoui, J.~Fatome, J.~C. Jules, B.~Kibler, G.~Gadret, C.~Fortier,
  F.~Smektala, I.~Skripatchev, C.F. Polacchini, Y.~Messaddeq, J.~Troles,
  L.~Brilland, M.~Szpulak, and G.~Renversez.
\newblock Strong infrared spectral broadening in low-loss {As-S} chalcogenide
  suspended core microstructured optical fibers.
\newblock {\em Opt. Express}, 18(5):4547--4556, 2010.

\bibitem{Akhmediev93}
Nail Akhmediev, Adrian Ankiewicz, and Jose~Maria Soto-Crespo.
\newblock Does the nonlinear {S}chrödinger equation correctly describe beam
  propagation?
\newblock {\em Opt. Lett.}, 18(6):411, 1993.

\bibitem{chen97maxwell-equations-vector-nonlinear-schrodinger-equation}
Y.~Chen and J.~Atai.
\newblock Maxwell's equations and the vector nonlinear {S}chrödinger equation.
\newblock {\em Physical Review E}, 55:3652--3657, 1997.

\bibitem{drouart08beyond-townes-soliton}
F.~Drouart, G~.Renversez, A.~Nicolet, and C.~Geuzaine.
\newblock Spatial {K}err solitons in optical fibers of finite size cross
  section: beyond the {T}ownes soliton.
\newblock {\em J. Opt. A: Pure Appl. Opt.}, 10:125101, 2008.

\bibitem{these-wiktor14}
W.~Walasik.
\newblock {\em Plasmon–soliton waves in metal-nonlinear dielectric planar
  structures}.
\newblock phdthesis, {A}ix-{M}arseille {U}niversité, 2014.

\bibitem{taflove13}
A.~Taflove, A.~Oskooi, and S.~G. Johnson, editors.
\newblock {\em Advances in {FDTD} Computational Electrodynamics}.
\newblock Photonics and Nanotechnology. Artech House, Boston, 2013.

\bibitem{Falconieri1999}
M.~Falconieri and G.~Salvetti.
\newblock Simultaneous measurement of pure-optical and thermo-optical
  nonlinearities induced by high-repetition-rate, femtosecond laser pulses:
  application to cs2.
\newblock {\em Applied Physics B}, 69(2):133--136, Aug 1999.

\bibitem{Gnoli05}
Andrea Gnoli, Luca Razzari, and Marcofabio Righini.
\newblock Z-scan measurements using high repetition rate lasers: how to manage
  thermal effects.
\newblock {\em Opt. Express}, 13(20):7976--7981, Oct 2005.

\end{thebibliography}
}
}{}
\end{document}